\documentclass[letterpaper,12pt]{article}
\usepackage[latin1]{inputenc}
\usepackage{amssymb}

\usepackage{ifpdf}
\ifpdf
	\usepackage[pdftex]{graphicx}
	\usepackage[pdftex,unicode,implicit]{hyperref}

	\hypersetup{
  	pdftitle     = {Supersymmetric solutions of gauged five-dimensional 		
  			supergravity with general matter couplings},
  	pdfkeywords  = {Supersymmetry, gauged supergravity, real special geometry, 
     	              BPS solutions},
  	pdfauthor    = {J. Bellor\'{\i}n},
  	pdfcreator   = {pdf\LaTeXe\ with package \flqq hyperref\frqq},
  	pdfproducer  = {pdf\LaTeXe\ with package \flqq hyperref\frqq},
  	pdfpagemode  = UseNone,  
  	pdffitwindow = true,  
  	unicode      = true,
  	plainpages   = true,
  	colorlinks   = true,  
  	citecolor    = blue,  
  	urlcolor     = blue,
  	linkcolor    = blue
	}

	\newcommand{\hepth}[1]{
		arXiv:{\tt\href{http://www.arXiv.org/abs/hep-th/#1}{hep-th/#1}}}

	\newcommand{\arxiv}[1]{
		{\tt\href{http://www.arXiv.org/abs/#1}{arXiv:#1}}}

\else
  \usepackage[dvips]{graphicx}
  \usepackage[unicode,implicit]{hyperref}
  \newcommand{\hepth}[1]{arXiv:{\tt hep-th/#1}}

  \newcommand{\arxiv}[1]{{\tt arXiv:#1}}
  
\fi

\makeatletter
\@addtoreset{equation}{section}
\makeatother

\begin{document}

\begin{flushright}
{\small
SB/F/363-08\\
Apr $15^{\rm th}$, $2009$}
\end{flushright}

\begin{center}

\vspace{2cm}
{\LARGE {\bf 
Supersymmetric solutions of gauged five-dimensional supergravity with general matter couplings}} 
\vspace{2cm}

{\sl\large Jorge Bellor\'{\i}n}

\vspace{1cm}

{\it Departamento de F\'{\i}sica, Universidad Sim\'on Bol\'{\i}var, Valle de Sartenejas,\\ 
1080-A Caracas, Venezuela.} \\
{\tt jorgebellorin@usb.ve} \\

\vspace*{2cm}
{\bf Abstract}

\end{center}

\begin{quotation}\small
We perform the characterization program for the supersymmetric configurations and solutions of the $\mathcal{N}=1$, $d=5$ Supergravity Theory coupled to an arbitrary number of vectors, tensors and hypermultiplets and with general non-Abelian gaugins. By using the conditions yielded by the characterization program, new exact supersymmetric solutions are found in the $SO(4,1)/SO(4)$ model for the hyperscalars and with $SU(2)\times U(1)$ as the gauge group. The solutions contain also non-trivial vector and massive tensor fields, the latter being charged under the $U(1)$ sector of the gauge group and with selfdual spatial components. These solutions are black holes with $AdS_2 \times S^3$ near horizon geometry in the gauged version of the theory and for the ungauged case we found naked singularities. We also analyze supersymmetric solutions with only the scalars $\phi^x$ of the vector/tensor multiplets and the metric as the non-trivial fields. We find that only in the null class the scalars $\phi^x$ can be non-constant and for the case of constant $\phi^x$ we refine the classification in terms of the contributions to the scalar potential.
\end{quotation}

\newpage

\pagestyle{plain}




\section{Introduction}

The seminal work on the characterization of supersymmetric solutions of supergravity was done by Tod in Ref.~\cite{Tod:1983pm} for the pure $\mathcal{N}=2$, $d=4$ Supergravity Theory. The method he used is based on the translation of the Killing spinor equations to tensorial equations for bilinears of the Killing spinors, which are easier to handle that the spinorial equations. All the conditions needed to have a bosonic supersymmetric configuration can be extracted from these equations for the bilinears. Furthermore, the Killing spinor equations are first-order, unlike the equations of motion that are second-order. Indeed, among the conditions found by Tod, it is remarkable the simplicity of the differential equations that must be satisfied by the supersymmetric solutions. Some years after this work, Tod extended the analysis to more general four dimensional supergravity theories in Ref.~\cite{Tod:1995jf}.

After the work of Tod, Gauntlett \emph{et al.} in Ref.~\cite{Gauntlett:2002nw} adapted the program to the minimal five-dimensional ungauged theory with eight supercharges ($\mathcal{N}=1$). These authors developed the approach in a pure tensorial language, in contrast to the Newmann-Penrose formalism used by Tod.

The work of Gauntlett \emph{et al.} renewed the attention on the characterization program and after it a lot of related work has been done for supergravity theories in several dimensions, number of supercharges, gauged/ungauged symmetries and matter couplings. Regarding the $\mathcal{N}=1$, $d=5$ theory, soon after the publication of Ref.~\cite{Gauntlett:2002nw} the classification in the gauged minimal theory was achieved in Ref.~\cite{Gauntlett:2003fk}. The analysis of the (Abelian) gauged theory with couplings to an arbitrary number of vector multiplets was done in Ref.~\cite{Gutowski:2004yv}, restricted to the time-like case and considering only symmetric spaces for the scalar manifold. The analysis was extended to the null case and relaxing the condition of symmetry on the scalar manifold in Ref.~\cite{Gutowski:2005id}\footnote{Previous work on these theories can be found in Refs.~\cite{Chamseddine:1998yv,Sabra:1997yd}.}. The characterization program for the ungauged theory with vector- and hypermultiplets (without restrictions on the scalar manifolds) was done in Ref.~\cite{Bellorin:2006yr} and the case of the theory with general non-Abelian gaugings was analyzed in Ref.~\cite{Bellorin:2007yp} (previous, partial analyzes were done in Refs.~\cite{Celi}).

As we mentioned above, there are many papers \cite{variousworks,{Jong:2006za}} on the characterization of supersymmetric solutions in others theories of supergravity. In addition, various papers \cite{spinorialgeometry} have been written on the subject using the technique of the spinorial geometry.

The main aim of this work is to complete the characterization of supersymmetric solutions of the $\mathcal{N}=1$, $d=5$ Supergravity Theory, initiated in Refs.~\cite{Gauntlett:2002nw, Gauntlett:2003fk, Gutowski:2004yv, Gutowski:2005id, Bellorin:2006yr, Bellorin:2007yp}, by turning on tensor multiplets. With the adding of tensor multiplets we cover the most general couplings to matter multiplets known in $\mathcal{N}=1$, $d=5$ Supergravity. We perform the study for the gauged version of the theory because it is a more general context and, most importantly, tensor multiplets make sense only in the gauged scenario. Any result in the ungauged theory can be obtained from the gauged theory by sending the Yang-Mills coupling constant to zero and turning off the tensor fields.

Tensor multiplets contain two-form fields that, in contrast to the field strengths of vector gauge fields, transform in a non-adjoint representation of the gauge group \cite{Gunaydin:1999zx}. Their exterior covariant derivative does not vanish,  they satisfy instead a massive self-duality condition \cite{Townsend:1983xs} which is precisely given by their equation of motion. This condition is needed in order to balance the fermionic/bosonic degrees of freedom in each multiplet, hence tensor multiplets must be considered as \emph{on-shell} multiplets\footnote{Of course, after all we are interested in supersymmetric solutions, hence all fields are on-shell for us.}. In addition, as well as vector multiplets, each tensor multiplet includes an scalar field and an one-half spinor field. 

The $\mathcal{N}=1$, $d=5$ gauged Supergravity coupled to an arbitrary number of vector-, tensor- and hypermultiplets is the most general theory of supergravity known with eight supercharges in five dimensions. Therefore, the results of the characterization can be used for very general analyses related to supersymmetric solutions, since preserved supersymmetry gives a lot of information about the solutions and the set of equations that must be solved is much smaller than the original set of equations of motion. For example, the conditions for supersymmetric solutions resulting from the characterization program can be used to analyze five-dimensional black holes/rings, which is a growing area of research.

One of the main interest on supersymmetric solutions of the five-dimensional supergravity is the existence of supersymmetric domain walls that have $AdS_5$ asymptotics. This kind of configurations is crucial for the brane-world scenario and also for the AdS/CFT correspondence. Their existence is related to the extrema of the scalar potential characteristic of the gauged theory. Tensor multiplets, as well as vector- and hypermultiplets, give contributions to this potential. Therefore, the possibilities of finding new classes of supersymmetric minima of the potential are increased once the characterization of supersymmetric solutions with the most general matter couplings is done. With the aim to get some insight on this class of configurations, in this paper we analyze supersymmetric scalar-gravity configurations once the characterization program has been completed. These configurations have vanishing vector and tensor fields and the hyperscalar are just constant, hence the scalars of the vector/tensor multiplets and the metric are the only active fields. Scalar-gravity configurations are close to pure gravity configurations and are appropriated to study minima of the scalar potential.

In order to give a further proof of the usefulness of the characterization of the supersymmetric solutions, in this paper we use the results of this program as a framework to find new supersymmetric solutions of $\mathcal{N} = 1$, $d=5$ Supergravity with several active matter fields. In Ref.~\cite{Bellorin:2006yr} new exact supersymmetric solutions of the ungauged $\mathcal{N} = 1$, $d=5$ supergravity theory with hyperscalars taking their value in the $SO(4,1)/SO(4)$ manifold were found. 
The setting used was taken from a solution of the six-dimensional theory found in Ref.~\cite{Jong:2006za}. It is intersting to see wether these 5d solutions can be generalized to the gauged case and with more active matter fields. Moreover, The 5d solutions found in Ref.~\cite{Bellorin:2006yr} have a naked singularity whereas the 6d solution found in Ref.~\cite{Jong:2006za}, which belongs to the gauged theory, has a horizon. Therefore, it is also interesting to determine if the presence of naked singularities is linked to the ungauged scenario. Motivated from these previous results on the $SO(4,1)/SO(4)$ model for the hyperscalars, we take this model to find the new supersymmetric solutions. To this end it also required to set an appropriated model for the scalar fields of the vector/tensor multiplets, as well as the gauge group together with the representation under which the tensor fields transform.

This paper is organized as follows: In the next section we show the generalities of the theory, including the action, equations of motion and supersymmetry transformation rules. In section 3 we perform the characterization program. We subdivide this section in three parts: general results, time-like case and null case. In section \ref{sec:vacuum} we analyze the supersymmetric scalar-gravity configurations. In section \ref{sec:so41so4} we show exact supersymmetric solutions in the $SO(4,1)/SO(4)$ model for the hyperscalars. Finally we present some conclusions.


\section{$\mathcal{N}=1,d=5$ Supergravity with gaugings and general matter couplings}
\label{sec:theory}

The theory was elaborated by several authors in successive steps. The Lagrangians of the pure $\mathcal{N}=1$, $d=5$ Supergravity were found in Refs.~\cite{originaln1d5}. The couplings of the pure theory to vector multiplets in the ungauged and gauged scenarios were analyzed in Refs.~\cite{Gunaydin:1983bi,Gunaydin:1984ak} and the extension to tensor multiplets was done in Ref.~\cite{Gunaydin:1999zx}. The first complete Lagrangian of gauged $\mathcal{N}=1$, $d=5$ Supergravity coupled to vector, tensor and hypermultiplets with general scalar manifolds was given in Ref.~\cite{Ceresole:2000jd}. The authors of Refs.~\cite{Gunaydin:1999zx,Ceresole:2000jd}, however, considered only a restricted class of couplings between vector and tensor fields. The theory with the most general couplings was achieved in Ref.~\cite{Bergshoeff:2004kh}, and we take the action from this reference, following most of its conventions\footnote{\hspace{1ex}In contrast to Ref.~\cite{Bergshoeff:2004kh}, we use a mostly minus signature for the space-time metric and define the Ricci tensor by $R_{\mu\nu} = R_{\mu\alpha\nu}{}^{\alpha}$. We raise and lower the $SU(2)$ index according to $A_i = \epsilon_{ij} A^j$, $A^i = -\epsilon^{ij} A_j$. We also fix the value of the five-dimensional Newton constant, $\kappa^{-1} = \sqrt2$.}

The field content of the theory consists of the supergravity multiplet $\{e_\mu{}^a, \psi_\mu^i,A_\mu\}$, $n_V$ copies of a vector multiplet, $\{A_\mu^{x'}, \phi^{x'}, \lambda^{x'i}\}$, $x'=1,\ldots,n_V$, $n_T$ copies of a tensor multiplet, $\{B_{\mu\nu}^M, \phi^{M}, \lambda^{Mi}\}$, $M=1,\ldots,n_T$, and $n_H$ copies of a hypermultiplet, $\{q^X, \zeta^{iA}\}$, $X=1,\ldots,4n_H$. As it is well-known, several fields of the supergravity, vector and tensor multiples are related between them by symmetries hence it is convenient to treat them on the same footing. From now on we use the following notation: $\phi^{x}$ and $\lambda^{xi}$, $x=1,\ldots,n_V+n_T$, are the scalar and spinor fields of the vector-tensor multiplets; $A^{I}$, $I=1,\ldots,n_V+1$ are the vector fields, including the vector field of the supergravity multiplet, and $H^{\tilde{I}}$, $\tilde{I}=1,\ldots,n_V+n_T+1$ represents the field strengths/tensor fields, $H^{\tilde{I}} = (F^I, B^M)$. In some instances it is convenient to unify the notation on the scalars, hence we use sometimes the index $\tilde{x}$ and the variable $\varphi^{\tilde{x}} = (\phi^x , q^X)$. All the spinors satisfy a symplectic-Majorana condition.

The target manifold of the $\phi^x$ fields belongs to the class of very special geometry whereas the target of the $q^X$ fields is a non-compact quaternionic K\"ahler manifold. We denote these manifolds by $\mathcal{M}_{VS}$ and $\mathcal{M}_{QK}$ respectively. The former is defined through the embedding of $\phi^x$ into $h^{\tilde{I}}(\phi^x)$, such that $\mathcal{M}_{VS}$ is defined by
\begin{equation}
C_{\tilde{I}\tilde{J}\tilde{K}} h^{\tilde{I}} h^{\tilde{J}} h^{\tilde{K}} 
= 1 \,,
\label{eq:veryspecial}
\end{equation}
where $C_{\tilde{I}\tilde{J}\tilde{K}}$ is a completely symmetric constant symbol. Given $C_{\tilde{I}\tilde{J}\tilde{K}}$ one has the scalar metrics
\begin{equation}
a_{\tilde{I}\tilde{J}} = 
- 2 C_{\tilde{I}\tilde{J}\tilde{K}} h^{\tilde{K}}
+3h_{\tilde{I}} h_{\tilde{J}} \,,
\hspace{2em}
g_{xy} = h^{\tilde{I}}_x h^{\tilde{J}}_y a_{\tilde{I}\tilde{J}} \,,
\label{eq:specialmetrics}
\end{equation}
where $h_{\tilde{I}} =  
C_{\tilde{I}\tilde{J}\tilde{K}} h^{\tilde{J}} h^{\tilde{K}}$ and $h^{\tilde{J}}_{x} = -\sqrt{3} \partial_ x h^{\tilde{J}}$, $h_{\tilde{J}x} = +\sqrt{3} \partial_x h_{\tilde{J}}$. Various identities of the very special geometry can be found in Ref.~\cite{Gunaydin:1983bi}

Quaternionic K\"ahler manifolds are $4n_H$-dimensional Riemannian manifolds characterized by a metric $g_{XY}$ and a quaternionic structure $\vec{J}_X{}^Y$, such that the three two-forms $\vec{J}_{XY}$ are covariantly closed respect to a $\mathfrak{su}(2)$ connection $\vec{\omega}_X$ whose curvature is proportional to $\vec{J}_{XY}$. In general, the holonomy of a quaternionic K\"ahler manifold is $SU(2)\times Sp(n_H,\mathbb{H})$. For $n_H=1$, however, these conditions are not restrictive at all. In this case quaternionic K\"ahler manifolds are defined by the conditions of being Einstein and self-dual. The subject of quaternionic-like manifolds is treated in detail in Ref.~\cite{Bergshoeff:2002qk}.

In the \emph{gauged} theory the vector fields are the gauge fields for some group $G$ that also acts on the ``matter" fields of the theory (scalars, tensors and spinors), which leads to couplings between gauge and matter fields via covariant derivatives\footnote{In the ungauged theory the vector fields have still Abelian gauge symmetries, but they do not act on the matter fields. In both cases there are other couplings between gauge and matter fields besides the covariant derivatives.}. We shall use names as ``G-symmetry'' to refer to this kind of gauge transformations. 

Since the Lagrangian is a $\sigma$-model for the scalars, the $G$-group must be a subgroup of the isometries of the target manifold. Therefore the directions of the gaugins are determined by Killing vectors satisfying
\begin{equation}
[k_I, k_J]^{\tilde{x}} = -f_{IJ}{}^K k_K{}^{\tilde{x}} \,.
\end{equation}
To each Killing vector of the $\mathcal{M}_{QK}$ manifold there is associated a triplet of scalars that form a $SU(2)$ vector,
\begin{equation}
2n_H\vec{P}_I = \vec{J}_X{}^Y \nabla_Y k_I{}^X \,,
\end{equation}	
which is called the momentum map. Momentum maps allow the embedding of the $G$-transformations into $SU(2)$ transformations, which is the R-symmetry group acting on the spinors. In particular, the corresponding $SU(2)$ gauge connection is $\frac{1}{2} A^I \vec{P}_I$. In addition, there are R-symmetry gauge transformations induced by $SU(2)$ transformations in $\mathcal{M}_{QK}$, such that the full space-time gauge connection for the R-symmetry is the combination 
\begin{equation}
\vec{B} 
= 
dq^X \vec{\omega}_X 
+{\textstyle\frac{1}{2}} g A^I \vec{P}_I \,.
\end{equation}
For instance, the covariant derivative on the gravitino is\footnote{We use the symbol $\mathfrak{D}$ to denote the generalized space-time covariant derivative. It is made from the affine, spin and vector gauge connections. When it acts on $SU(2)$ objects like the spinors it includes the $\vec{B}$ connection.}
\begin{equation}
\mathfrak{D}_\mu \psi_\nu^i = \nabla_\mu \psi_\nu^i + B_{\mu j}{}^i \psi^j_\nu \,.
\end{equation}
In absence of hypermultiplets ($n_{H}=0$) the momentum maps $\vec{P}_{I}$ can
still be defined in two cases in which they are equivalent to a set of
constant Fayet-Iliopoulos terms. In the first case the gauge group contains
an $SU(2)$ factor and
\begin{equation}
\vec{P}_{I} = \vec{e}_{I}\,\xi\, ,
\end{equation}
where $\xi$ is an arbitrary constant and the $\vec{e}_{I}$ are constants that
are nonzero for $I$ in the range of the $SU(2)$ factor and satisfy
\begin{equation}
\vec{e}_{I}\times \vec{e}_{J} = f_{IJ}{}^{K}\vec{e}_{K}\, .  
\end{equation}
In the second case the gauge group contains $U(1)$ factors and 
\begin{equation}
\vec{P}_{I} = \vec{e}\,\xi_{I}\, ,
\end{equation}
where $\vec{e}$ is an arbitrary $SU(2)$ vector and the $\xi_{I}$s are
arbitrary constants that are nonzero for the $I$ corresponding to $U(1)$
factors.

Tensor multiplets can be coupled only in the gauged version of the theory. In general, the vector/tensor multiplets are mixed under the action of the $G$-group. They transform in a representation whose basis is denoted by $t_I$,
\begin{equation}
[t_I,t_J] = -f_{IJ}{}^K t_K \,.
\label{eq:galgebra}
\end{equation}
The general form of these matrices is
\begin{equation}
[t_I]_{\tilde J}{}^{\tilde K} =
\left(\begin{array}{cc}
f_{IJ}{}^K & t_{IJ}{}^M \\
0 & t_{IN}{}^M
\end{array}
\right)\,,
\end{equation}
where, of course, the adjoint-representation components correspond to the gauge sector, whereas the rest of components are not in the adjoint representation and correspond to the transformations of the tensor multiplets. Although $f_{IJ}{}^K$ and $t_{IN}{}^M$ realize smaller representations of the $G$-group, the presence of the $t_{IJ}{}^M$ components makes $t_{I\tilde{J}}{}^{\tilde{K}}$ a non completely reducible representation. It is known \cite{Gunaydin:1999zx} that the supersymmetry of the theory forbids the coupling to vector fields that are ``charged'' under the $G$-gauge group (i. e., transforming homogeneously in a representation that is not the adjoint), since this kind of vector fields would be massive and this leads to a mismatch between the fermionic/bosonic degrees of freedom. Massive tensor fields are the only admissible charged two-forms in the theory. 

The Killing vectors of the target manifold $\mathcal{M}_{VS}$ are completely determined in terms of the matrices $t_I$,
\begin{equation}
k_I{}^x = -\sqrt{3} t_{I\tilde J}{}^{\tilde K} h^{\tilde Jx} h_{\tilde K}
= -\sqrt{3} t_{I\tilde J}{}^{\tilde K} h^{\tilde J} h_{\tilde K}^x \,.
\label{eq:killingspecial}
\end{equation}

The $G$-symmetry transformations and covariant derivatives of the bosonic fields are
\begin{eqnarray}
&&
\delta_\Lambda A^I = 
d\Lambda^I + g f_{JK}{}^I A^J \Lambda^K \,, 
\hspace{2em}
\delta_\Lambda X^{\tilde I} = 
- g \Lambda^J t_{J\tilde K}{}^{\tilde I} X^{\tilde K}\,,
\\ \nonumber \\
&& 
\delta_\Lambda \varphi^{\tilde{x}} = -g \Lambda^I k_I^{\tilde{x}} \,,
\\ \nonumber \\
&&
\mathfrak{D} X^{\tilde{I}} = 
dX^{\tilde{I}} 
+ g A^{J} t_{J\tilde{K}}{}^{\tilde{I}} X^{\tilde{K}} \,,
\hspace{2em}
\mathfrak{D}\varphi^{\tilde{x}}=
d\varphi^{\tilde{x}} + g A^I k_I{}^{\tilde{x}} \,,
\end{eqnarray}
where $X^{\tilde I}$ can be $H^{\tilde I }$, $h^{\tilde I}$ and so on. 

The action of the theory needs one more symbol: an antisymmetric and invertible matrix $\Omega_{MN}$ (the number of tensor multiplets is restricted to be even).
We denote by $\Omega^{MN}$ the (minus) inverse of $\Omega_{MN}$, $\Omega_{MN} \Omega^{PN} = \delta_M{}^P$. Invariance of the action under $G$-symmetry transformations imposes the constraints
\begin{eqnarray}
&& t_{I[M}{}^P \Omega_{N]P} = 0 \,,
\label{eq:reducible}
\\ \nonumber \\
&& t_{I(\tilde J}{}^{\tilde F} C_{\tilde K\tilde L)\tilde F} = 0 \,.
\label{eq:Cgauge}
\end{eqnarray}
Moreover, it turns out that the tensorial components of $C_{\tilde{I}\tilde{J}\tilde{K}}$ are not independent parameters, they are instead given by\footnote{\hspace{1ex}$t_M =0$}
\begin{equation} 
C_{\tilde J\tilde K M} = 
-{\textstyle\frac{\sqrt{3}}{2}} t_{(\tilde J\tilde K)}{}^P \Omega_{PM} \,.
\label{eq:CM}
\end{equation}
Notice that, in order to avoid singular metrics $a_{\tilde{I}\tilde{J}}$ and $g_{xy}$ for a theory coupled to tensor multiplets, we must demand that not all the components $t_{I\tilde{J}}{}^M$ vanish, that is, tensor multiplets can be coupled only in the gauged theory and only when they are charged under some sector of the $G$-group.

The bosonic action of $\mathcal{N}=1,d=5$ gauged supergravity is given by\footnote{\hspace{1ex} In this formula we omit the wedge symbol in the multiplication of differential forms}
\begin{equation}
\begin{array}{rcl}
S & = &  {\displaystyle\int} \Bigl[R\star 1
+{\textstyle\frac{1}{2}}g_{\tilde x \tilde y} \mathfrak{D}\varphi^{\tilde x} \star\mathfrak{D}\varphi^{\tilde y}
+\mathcal{V}(\phi,q) \star 1
-{\textstyle\frac{1}{2}} a_{\tilde I\tilde J} 
H^{\tilde I} \star H^{\tilde J}
\\ \\ & &
+{\textstyle\frac{1}{3\sqrt{3}}}C_{IJK} F^{I} F^{J} A^{K}  -{\textstyle\frac{1}{4}} g S_{IJKL} F^I A^J A^K A^L  
+{\textstyle\frac{1}{40}} g^2 S_{IJKL} f_{FG}{}^I A^J A^K A^L A^F A^G
\\ \\
& &
+{\textstyle\frac{1}{4}}\Omega_{MN} \left(
g^{-1}B^M dB^N + 2 t_{IJ}{}^N B^M A^I F^J
+ t_{IP}{}^N B^M A^I  B^P \right)
\Bigr] \,,
\end{array}
\end{equation}
where 
\begin{eqnarray}
\mathcal{V}(\phi,q) 
&=& 
4g^2 \left( 2 \vec{P} \cdot \vec{P}
  - \vec{P}_x \cdot \vec{P}^x
  -{\textstyle\frac{1}{8}} k^{\tilde x} k_{\tilde x}\right) \,,
\label{eq:scalarpotential}
\\ \nonumber \\
& = &
4g^{2} \left( C^{IJK} h_{I} \vec{P}_J \cdot \vec{P}_K
-{\textstyle\frac{1}{8}} k^{\tilde x} k_{\tilde x}\right)\, ,
\\ \nonumber \\
\vec{P} &=&
{\textstyle\frac{1}{\sqrt{2}}} h^I \vec{P}_I \,, 
\hspace*{2em} 
\vec{P}_x \ = \
{\textstyle\frac{1}{\sqrt{2}}} h^I_x \vec{P}_I \,, 
\hspace*{2em} 
k^{\tilde{x}} \ = \ 
\sqrt{3} h^I k_I^{\tilde{x}} \,,
\\ \nonumber \\
S_{IJKL} &=& 
{\textstyle\frac{1}{3\sqrt3}} C_{FIJ} f_{KL}{}^F
-{\textstyle\frac{1}{2}} \Omega_{MN} t_{JI}{}^M t_{KL}{}^N \,.
\end{eqnarray}

The equations of motion of the bosonic fields, for which we use the following notation
\begin{equation}
\mathcal{E}_a{}^\mu \equiv 
-\frac{1}{2\sqrt{g}} \frac{\delta S}{\delta e^a{}_\mu} \,, \;\;
\mathcal{E}_{\tilde x} \equiv 
-\frac{1}{\sqrt{g}} \frac{\delta S}{\delta \varphi^{\tilde x}} \,, \;\;
\mathcal{E}_I{}^\mu \equiv 
\frac{1}{\sqrt{g}} \frac{\delta S}{\delta A^I{}_\mu} \,, \;\;
\mathcal{E}_M{}^{\mu\nu} \equiv 
\frac{4}{\sqrt{g}} \frac{\delta S}{\delta B^M{}_{\mu\nu}} \,, 
\end{equation}
are 
\begin{eqnarray}
\mathcal{E}_{\mu\nu} 
& = &   
G_{\mu\nu}
-{\textstyle\frac{1}{2}}a_{\tilde I\tilde J}\left(H^{\tilde I}{}_{\mu}{}^{\rho} H^{\tilde J}{}_{\nu\rho}
-{\textstyle\frac{1}{4}}g_{\mu\nu}
H^{\tilde I\, \rho\sigma}H^{\tilde J}{}_{\rho\sigma}\right)      
+{\textstyle\frac{1}{2}}g_{\tilde x \tilde y}\left(
\mathfrak{D}_{\mu}\varphi^{\tilde x} \mathfrak{D}_{\nu}\varphi^{\tilde y}
-{\textstyle\frac{1}{2}}g_{\mu\nu}
\mathfrak{D}_\rho\varphi^{\tilde x} \mathfrak{D}^{\rho}\varphi^{\tilde y} \right)
\nonumber \\ \nonumber \\ & &
-{\textstyle\frac{1}{2}}g_{\mu\nu}\mathcal{V}\, ,
\label{eq:Emn} \\
& & \nonumber \\
\mathcal{E}^{x} 
& = & 
\mathfrak{D}_{\mu}\mathfrak{D}^{\mu}\phi^{x} 
+{\textstyle\frac{1}{4}} \partial^x
a_{\tilde I\tilde J} H^{\tilde I\, \rho\sigma}H^{\tilde J}{}_{\rho\sigma}
-\partial^x\mathcal{V} \,,
\label{eq:Ei} 
\\ & & \nonumber \\
\mathcal{E}^X &=&
\mathfrak{D}_\mu \mathfrak{D}^\mu q^X - \partial^X \mathcal{V} \,,
\label{eq:EX}
\\ \nonumber \\
\star \mathcal{E}_{I} 
& = & 
-d \star H_I
+g A^J f_{JI}{}^K \star H_K
+{\textstyle\frac{1}{2}} A^J t_{JI}{}^M \Omega_{MN} \mathfrak{D}B^N
+{\textstyle\frac{1}{\sqrt{3}}} 
C_{I\tilde J\tilde K} H^{\tilde J} H^{\tilde K}
+g k_{I\tilde x} \star \mathfrak{D} \varphi^{\tilde x}  \, ,
\nonumber \\
\label{eq:preERm}
\\ \nonumber \\
\star {\cal E}_M & = &
g^{-1} \Omega_{MN} \mathfrak{D} B^N - 2\star H_M \,. 
\label{eq:EM}
\end{eqnarray}

Note that the Maxwell equation (\ref{eq:preERm}) is $G$-gauge-invariant only when the $\mathcal{E}_M$ equation is strictly on-shell. However, we may define the alternative equation
\begin{eqnarray}
\star\tilde\mathcal{E}_I &\equiv& 
\star\mathcal{E}_I
-{\textstyle\frac{1}{2}} g A^J t_{JI}{}^M  \star\mathcal{E}_M \,,
\\ \nonumber \\
&=&
-\mathfrak{D}\star H_I 
+ {\textstyle\frac{1}{\sqrt3}}C_{I\tilde J\tilde K}H^{\tilde J} H^{\tilde K}
+gk_{I\tilde x}\star\mathfrak{D}\varphi^{\tilde x} \,,
\label{eq:ERm}
\end{eqnarray}
whose gauge transformation is
\begin{equation}
\delta_\Lambda \star \tilde\mathcal{E}_I =
g\Lambda^J f_{JI}{}^K \star\tilde{\mathcal{E}}_K
+g \Lambda^J t_{JI}{}^M ({\textstyle\frac{1}{2}}\mathfrak{D}\star \mathcal{E}_M)
\,.
\end{equation}
Therefore, the pair $(\tilde\mathcal{E}_I^{\mu}, -\frac{1}{2}\mathfrak{D}_\nu\mathcal{E}_M^{\nu\mu})$ is a doublet of the $G$-symmetry group. When one deals with solutions, which requires the vanishing of all equation of motion, one may freely choose between the Maxwell equation (\ref{eq:preERm}) or the alternative equation (\ref{eq:ERm}). The use of Eq.~(\ref{eq:ERm}) has the advantage that the $G$-symmetry-covariance is kept manifest \emph{even off-shell}. This property is particularly useful for the analysis of the Killing Spinor Identities \cite{Kallosh:1993wx,Bellorin:2005hy}, which we are going to study in detail in section \ref{sec:susy}. In the KSIs one regards all the equations of motion as off-shell expressions in order to determine which of them are really independent when they are evaluated on supersymmetric configurations. 

Eq.~(\ref{eq:EM}) can be considered as a massive selfduality condition for two-forms in five dimensions \cite{Townsend:1983xs} and, moreover, it is the integral of the Proca equation. To see this, consider for a moment the Eq.~(\ref{eq:EM}) evaluated on vanishing v.e.v. for scalars and vectors,
\begin{equation}
dB^M - 2 g \Omega^{MN} \star B^N = 0 \,,
\end{equation}
where we have assumed $a_{\tilde{I}\tilde{J}} = \delta_{\tilde{I}\tilde{J}}$. Operating this equation with $\star d \star$ we get
\begin{equation}
 \star d \star d B^M + [\mathcal{M}^2]^M{}_N B^N = 0 \,,
\label{eq:proca}
\end{equation}
where 
\begin{equation}
 [\mathcal{M}^2]^M{}_N = - 4 g^2 \Omega^{MP} \Omega^{PN}  \,.
\label{eq:tensormass}
\end{equation}
Eq.~(\ref{eq:proca}) is the Proca equation for a two-form. As usual, this equation is equivalent to $d\star B^M = 0$ and the Klein-Gordon equation $(\delta^M{}_N \nabla^2 + [\mathcal{M}^2]^M{}_N ) B^M = 0$. Therefore the eigenvalues of $\mathcal{M}^2$ are the mass parameters for the tensor fields. 

The supersymmetry transformation rules for the fermionic fields, evaluated on
vanishing fermions, are
\begin{eqnarray}
\delta_{\epsilon}\psi^{i}_{\mu} 
& = & 
\mathfrak{D}_{\mu}\epsilon^{i}
-{\textstyle\frac{1}{8\sqrt{3}}}h_{\tilde I}H^{\tilde I\,\alpha\beta}
\left(\gamma_{\mu\alpha\beta}-4g_{\mu\alpha}\gamma_\beta\right)
\epsilon^{i}
+{\textstyle\frac{1}{\sqrt{6}}} g P_{j}{}^{i} \gamma_\mu\epsilon^j \, , 
\\
& & \nonumber \\ 
\delta_{\epsilon}\lambda^{ix} 
& = &  
{\textstyle\frac{1}{2}}\left(\not\!\!\mathfrak{D}\phi^{x} 
-{\textstyle\frac{1}{2}}h^{x}_{\tilde I}\not\!\!H^{\tilde I}
+ g k^x \right)\epsilon^{i}
+ \sqrt{2} g P^x{}_{j}{}^{i} \epsilon^{j} \, , 
\\
& & \nonumber \\ 
\delta_{\epsilon}\zeta^A 
& = &
{\textstyle\frac{1}{2}}\left( \not\!\!\mathfrak{D} q^X 
+ g k^X \right) f_X{}^{iA} \epsilon_i 
\end{eqnarray}
and the supersymmetry transformation rules of the bosonic fields are 
\begin{eqnarray}
\label{eq:susytranse}
\delta_{\epsilon} e^{a}{}_{\mu} 
& = & 
{\textstyle\frac{i}{2}} \bar{\epsilon}_{i}\gamma^a\psi^{i}_{\mu}\, ,
\\
& & \nonumber \\ 
\label{eq:susytransA}
\delta_{\epsilon} A^{I}{}_{\mu} 
& = & 
\vartheta^I_\mu\, ,
\\
& & \nonumber \\
\label{eq:susytransZ}
\delta_{\epsilon} \phi^{x} 
& = & 
{\textstyle\frac{i}{2}}\bar{\epsilon}_{i}\lambda^{x\,i}\,,
\\ \nonumber \\
\delta_\epsilon B^M_{\mu\nu} & = &
2\mathfrak{D}_{[\mu} \vartheta^M{}_{\nu]} 
+2\sqrt3 i g \bar\epsilon_i \gamma_{[\mu} \psi_{\nu]}^i h_N \Omega^{MN} 
+ig\bar\epsilon_i \gamma_{\mu\nu} \lambda^{xi} h_{Nx}\Omega^{MN}\,,
\label{eq:susytransB}
\\ \nonumber \\
\delta_\epsilon q^X & = &
-i\bar{\epsilon}^i \zeta^A f_{iA}{}^X
\label{eq:susytransq}
\,,
\end{eqnarray}
where
\begin{equation}
\vartheta^{\tilde I}_\mu  = 
-{\textstyle\frac{\sqrt3}{2}} ih^{\tilde I} \bar\epsilon_i\psi_\mu^i
+{\textstyle\frac{i}{2}}\bar\epsilon_i\gamma_\mu\lambda^{xi} h_x^{\tilde I} \,.
\end{equation}

Let us make a brief comment about pure gravity solutions. If we turn off vector and tensor fields, $A^I = B^M = 0$, the scalar (\ref{eq:Ei})-(\ref{eq:EX}) and Maxwell equations (\ref{eq:ERm}) become respectively
\begin{equation}
\nabla^2 \varphi^{\tilde{x}} 
- \partial^{\tilde{x}} \mathcal{V} = 0 \,,
\hspace*{2em}
 k_{I\tilde{x}} \partial_{\mu} \varphi^{\tilde{x}}  =  0 \,.
\end{equation}
The simplest solution to the last equation is $\partial_\mu \phi^x = \partial_\mu q^X = 0$, which implies for the former equation that the solution is a critical point of the potential, as expected for a pure gravity solution. If the scalars fields are constant all the objects that belongs to the target manifold, like $h^I$, $h^I_x$, $k_I^{\tilde{x}}$ and $\vec{P}_I$, are also constant as well as the scalar potential is (the cosmological constant). We shall find this kind of configurations in section \ref{sec:vacuum} as part of the supersymmetric solutions.


\section{Supersymmetric configurations and solutions}
\label{sec:susy}

We follow the procedure of Refs.~\cite{Bellorin:2006yr,Bellorin:2007yp}, where the standard programme based on the spinor bilinears was used to solve the KSEs combined with the computation of the KSIs in order to determine the independent supersymmetric equation of motion. In these papers the supersymmetric configurations and solutions of the ungauged/gauged $\mathcal{N}=1$, $d=5$ Supergravity coupled to vector- and hypermultiplets were characterized. Here we follow the same steps now including tensor multiplets, writting only the main steps.

\subsection{General results}

\subsubsection{Killing Spinor Equations and bilinears}
The KSEs are
\begin{eqnarray}
\label{gravitinokse}
\mathfrak{D}_{\mu}\epsilon^{i}
-{\textstyle\frac{1}{8\sqrt{3}}}h_{\tilde I}H^{\tilde I\,\alpha\beta}
\left(\gamma_{\mu\alpha\beta}-4g_{\mu\alpha}\gamma_\beta\right)\epsilon^{i} 
+{\textstyle\frac{1}{\sqrt{6}}} g P_{j}{}^{i} \gamma_\mu\epsilon^j 
& = & 0\, , 
\\ \nonumber \\
\label{gauginokse} 
\left(\not\!\!\mathfrak{D}\phi^{x} 
+ g k^x 
-{\textstyle\frac{1}{2}}h^{x}_{\tilde I}\not\!\!H^{\tilde I}
\right)\epsilon^{i} 
+ 2\sqrt{2} g  P^{x}{}_{j}{}^{i} \epsilon^{j}
& = & 0\, ,
\\ \nonumber \\
\left( \not\!\!\mathfrak{D} q^X 
+ g k^X \right) f_X{}^{iA} \epsilon_i
& = & 0 \,.
\label{eq:hyperinokse}
\end{eqnarray}
By comparing with the theory without tensor multiplets, we see that there is a new term in the gaugino KSE (\ref{gauginokse}) proportional to the projection $k^x \sim h^I k_I^x$. This projection automatically vanishes if there are no tensor multiplets. When tensor multiplets are turned on, it can be easily shown that
\begin{equation}
h^I k_I^x = 
- 2 \Omega^{MN} h_M h_N^x \,.
\label{eq:hk}
\end{equation}
The spinor bilinears that can be constructed from the Killing spinor are the scalar $f$, the vector $V$ and the three 2-forms $\Phi^{r}$. 

The corresponding differential equations for the bilinears are
\begin{eqnarray}
\label{df}df & = & {\textstyle\frac{1}{\sqrt{3}}}h_{\tilde I}i_{V}H^{\tilde I}\, , \\
& & \nonumber \\
\label{killingvector}                
\nabla_{(\mu}V_{\nu)} & = & 0\, ,\\
& & \nonumber \\
\label{dV}
dV & = & 
-{\textstyle\frac{2}{\sqrt{3}}}fh_{\tilde I}H^{\tilde I}
-{\textstyle\frac{1}{\sqrt{3}}}\star(h_{\tilde I}H^{\tilde I}\wedge V)
-{\textstyle\frac{2\sqrt{2}}{\sqrt{3}}}g \vec{P}\cdot\vec{\Phi}
\, ,\\
& & \nonumber \\
\mathfrak{D}_{\alpha}\vec{\Phi}_{\beta\gamma} & = &
-{\textstyle\frac{1}{\sqrt{3}}}h_{\tilde I}H^{\tilde I\,\rho\sigma}
(g_{\rho[\beta}\star \vec{\Phi}_{\gamma]\alpha\sigma}
-g_{\rho\alpha}\star\vec{\Phi}_{\beta\gamma\sigma} 
-{\textstyle\frac{1}{2}}g_{\alpha[\beta}\star
\vec{\Phi}_{\gamma]\rho\sigma})
\label{nablaPhi}
\nonumber \\ \nonumber \\ & &
+{\textstyle\frac{\sqrt{2}}{\sqrt{3}}}g\left[
 \vec{P} \times (\star\vec{\Phi})_{\alpha\beta\gamma}
+2g_{\alpha[\beta} V_{\gamma]} \vec{P}
\right]\,,
\end{eqnarray}
where 
\begin{equation}
\mathfrak{D}_{\alpha}\vec{\Phi}_{\beta\gamma}
=
\nabla_{\alpha}\vec{\Phi}_{\beta\gamma} 
+2 \vec{B}_{\alpha} \times \vec{\Phi}_{\beta\gamma}
\,,
\label{eq:defdphi}  
\end{equation}
and the algebraic ones are
\begin{eqnarray}
V^\mu\mathfrak{D}_\mu\varphi^{\tilde{x}} 
& = & 
- g f k^{\tilde{x}}\, ,
\label{lvphi}
\\ \nonumber \\ 
\label{preholomorphicq}
f\mathfrak{D}_{\mu} q^{X} +\Phi^{r}{}_{\mu}{}^{\nu}\mathfrak{D}_{\nu} 
q^{Y} J^{r}{}_{Y}{}^{X} & = & 
- g k^{X} V_{\mu} \, ,
\\ \nonumber \\
\label{dphi}
f\mathfrak{D}_\mu\phi^{x} - h^{x}_{\tilde I} H^{\tilde I}{}_{\mu\nu} V^\nu
& = & 
- g k^x V_\mu \, ,
\\ 
& & \nonumber \\ 
\label{eq:Phidphi}  
\vec{\Phi}_{\mu\nu} \mathfrak{D}^{\nu}\phi^{x}
+{\textstyle\frac{1}{4}}\epsilon_{\mu\nu\alpha\beta\gamma} 
h^{x}_{\tilde I} H^{\tilde I\,\nu\alpha}\vec{\Phi}^{\beta\gamma}
& = & 
- 2\sqrt{2} g \vec{P}^x V_{\mu} \, ,
\\ \nonumber \\
\label{FtimesPhi}
h^{x}_{\tilde I}H^{\tilde I}_{\alpha\beta}\vec{\Phi}^{\alpha\beta} 
& = & 
4\sqrt{2} g f \vec{P}^x
\, .
\end{eqnarray}
The differential equation for $\Phi^{r}$~(\ref{nablaPhi}) implies
\begin{equation}
\label{eq:covariantconstancy}
d\Phi^{r}+2\varepsilon^{rst}B^{s}
\wedge\Phi^{t}=
\sqrt{6} g \epsilon^{rst} P^{s}\:\star\Phi^{t}\,.
\end{equation}

Eq.~(\ref{killingvector}) says that $V$ is an isometry of the space-time
metric. We fix partially the $G$-symmetry using the condition
\begin{equation}
i_{V} A^{I} + \sqrt{3} fh^{I}=0\, .
\label{gaugefixing}
\end{equation}
In this gauge, the scalars $q^{X}$, $\phi^{x}$ and $f$ are independent of the coordinate adapted to the isometry $V$.


\subsubsection{Killing Spinor Identities}
\label{sec-ksis}
We use the Killing Spinor Identities (KSIs) \cite{Kallosh:1993wx,Bellorin:2005hy} as an easy way to determine which of the equations of motion are independent once they are evaluated on supersymmetric configurations. The KSIs are derived from the supersymmetry transformation rules of the bosons, which are listed in Eqs.~(\ref{eq:susytranse}) - (\ref{eq:susytransq}). 

The main difference we found respect to Refs.~\cite{Bellorin:2006yr,Bellorin:2007yp} is that, since now me have tensor multiplets, each of them containing scalar, tensor and spinor fields, there are new KSIs that arise as new components of the old KSIs that were obtained by taking derivatives respect to $\lambda^x$. Moreover, the supersymmetry transformation rule of the tensor fields is a totally new rule, hence the explicit form of the KSIs is modified by the presence of it. 

The KSIs that we obtain from Eqs.~(\ref{eq:susytranse}) - (\ref{eq:susytransq}) are
\begin{eqnarray}
\label{eq:preksi1}
 \left[
\mathcal{E}_{\mu\nu}\gamma^{\nu} 
+{\textstyle\frac{\sqrt{3}}{2}}(
h^{I}\tilde\mathcal{E}_{I\mu}
-{\textstyle\frac{1}{2}} h^M \mathfrak{D}^\nu {\cal E}_{M\nu\mu} )
-{\textstyle\frac{\sqrt{3}}{2}} g \Omega^{MN} h_M {\cal E}_{N\mu\nu} \gamma^\nu
\right]\epsilon^{i} 
& = & 
0\, ,\\
& & \nonumber\\
\label{eq:preksi2}
\left[
\mathcal{E}_{x} 
-(h^{I}_{x} \tilde\mathcal{E}_{I}^\mu 
-{\textstyle\frac{1}{2}} h^M_x \mathfrak{D}_\nu \mathcal{E}_M^{\nu\mu})\gamma_\mu
-{\textstyle\frac{1}{2}} g \Omega^{MN} h_{M\,x} \not\!{\cal E}_N
\right]\epsilon^{i}
& = & 
0\, ,
\\ \nonumber \\
f_{iA}{}^X \mathcal{E}_X \epsilon^i 
& = & 0 \,.
\end{eqnarray}
These are spinorial expressions. We may obtain the tensorial KSIs by contracting these equations with $(\sigma^m)_i{}^j\bar{\epsilon}_j$ and $(\sigma^m)_i{}^j\bar{\epsilon}_j \gamma_\alpha$, where $\sigma^m = (1,\vec{\sigma})$. The tensorial KSIs are
\begin{eqnarray}
\mathcal{E}_{\mu\nu} V^\nu 
+ {\textstyle\frac{\sqrt{3}}{2}} (h^I {\tilde\mathcal{E}}_I{}^\mu
    -{\textstyle\frac{1}{2}} h^M \mathfrak{D}^\nu \mathcal{E}_{M\nu\mu}) f
-{\textstyle\frac{\sqrt{3}}{2}} g \Omega^{MN} h_M \mathcal{E}_{N\mu\nu} V^\nu
& = & 0 \,,
\\ \nonumber \\
\mathcal{E}_x f 
-(h^I_x{\tilde\mathcal{E}}_{I\mu} - {\textstyle\frac{1}{2}} h^M_x \mathfrak{D}^\nu \mathcal{E}_{M\nu\mu}) V^\mu 
& = & 0 \,,
\\ \nonumber \\
g \Omega^{MN} h_{Mx} \mathcal{E}_N{}^{\mu\nu} \vec{\Phi}_{\mu\nu}
& = & 0 \,,
\\ \nonumber \\
\mathcal{E}_X f 
& = & 0 \,,
\\ \nonumber \\
\mathcal{E}_{\mu\nu} f
+ {\textstyle\frac{\sqrt{3}}{2}} (h^I {\tilde\mathcal{E}}_{I\mu}
    -{\textstyle\frac{1}{2}} h^M \mathfrak{D}^\alpha \mathcal{E}_{M\alpha\mu}) V_\nu
-{\textstyle\frac{\sqrt{3}}{2}} g \Omega^{MN} h_M \mathcal{E}_{N\mu\nu} f
& = & 0 \,,
\\ \nonumber \\
(\mathcal{E}_{\mu\nu} 
- {\textstyle\frac{\sqrt{3}}{2}} g \Omega^{MN} h_M \mathcal{E}_{N\mu\nu} ) \vec{\Phi}^{\nu}{}_\alpha
& = & 0 \,,
\\ \nonumber \\
\mathcal{E}_x V_\mu
-(h^I_x{\tilde\mathcal{E}}_{I\mu} - {\textstyle\frac{1}{2}} h^M_x \mathfrak{D}^\nu \mathcal{E}_{M\nu\mu}) f
+ g \Omega^{MN} h_{Mx} \mathcal{E}_{N\nu\mu} V^\nu
& = & 0 \,,
\\ \nonumber \\
(h^I_x {\tilde\mathcal{E}}_I{}^\mu 
  - {\textstyle\frac{1}{2}} h^M_x \mathfrak{D}_\nu \mathcal{E}_M{}^{\nu\mu})
\vec{\Phi}_{\mu\alpha} 
+{\textstyle\frac{1}{2}} g \Omega^{MN} h_{Mx} \mathcal{E}_{N\mu\nu} {\star\vec{\Phi}_\alpha{}^{\mu\nu}}
& = & 0 \,,
\\ \nonumber \\
\mathcal{E}_X V_\mu 
& = & 0 \,.
\end{eqnarray}

So far we have obtained the general equations and KSIs for the bilinears. To extract further information from these equations it is necessary to study separately the time-like ($f \neq 0$) and null ($f = 0$) cases. 


\subsection{The timelike case}
\label{sec-timelike}


\subsubsection{The equations for the bilinears}

The metric can be written in the form:
\begin{equation}
ds^{2} = f^{2}\left(dt+\omega\right)^{2}
-f^{-1}h_{\underline{m}\underline{n}} dx^{\underline{m}}dx^{\underline{n}}\, ,
\hspace{2em}
V = \partial_t
\label{conforma-stationary}
\end{equation}
with $\omega$ and $h_{\underline{m}\underline{n}}$ independent of time, as well as $f$ and $\varphi^{\tilde{x}}$ due to our partial $G$-gauge fixing (\ref{gaugefixing}).

The splitting of the gauge potential is
\begin{equation}
\label{eq:A} 
A^{I}  =  -\sqrt{3} h^{I} e^{0} + \hat{A}^{I} \,,
\end{equation}
where $e^0 = f(dt + \omega)$. The supersymmetric expression for $H^{\tilde{I}} = (F^I,B^M)$, obtained from the equations for the bilinears, is
\begin{equation}
H^{\tilde{I}} = -\sqrt{3}\ \hat{\mathfrak{D}}(h^{\tilde{I}}e^{0})+\hat{H}^{\tilde{I}} \, ,
\label{eq:FI}
\end{equation}
where $ \hat{\mathfrak{D}}$ is the 4-dimensional spatial covariant derivative with respect to $\hat{A}^{I}$, $\hat{H}^{\tilde{I}} = (\hat{F}^I,\hat{B}^M)$, $\hat{F}^{I}$ is the field strength of $\hat{A}^I$ and $\hat{B}^M$ are spatial two-forms. The magnetic component of these fields are subject to
\begin{eqnarray}
\label{remanentconstraint}
h_{\tilde I}\hat{H}^{\tilde I(+)} & = &  
{\textstyle\frac{2}{\sqrt{3}}} f (d\omega)^{(+)} \,,
\label{eq:domega}
\\ \nonumber \\
\hat{H}^{\tilde I(-)} & = & 
-2g f^{-1} C^{\tilde I\tilde J K} h_{\tilde J} \vec{P}_K \cdot \vec{\Phi} \,.
\label{eq:theta-}
\end{eqnarray}
These equations should be regarded as conditions on $\hat{A}^I$ instead of $\hat{F}^I$, otherwise we were forced to impose the corresponding Bianchi identity. The other point of view, however, could be fruitful in the ungauged case in which the gauge potential is not needed explicitly to build supersymmetric configurations.

Since there are not couplings between hyperscalars and tensor fields, the differential equation for the hyperscalars is of the same form of the case without tensor fields\footnote{From now on spatial flat indices refer to the 4-dimensional spatial metric $h_{\underline{mn}}$.}
\begin{equation}
\hat{\mathfrak{D}}_{m} q^{X} 
=
\Phi^{r}{}_{m}{}^{n} \hat{\mathfrak{D}}_{n} q^{Y} J^{r}{}_{Y}{}^{X} \,.
\label{triholomorphic}
\end{equation}

The projection of Eq.~(\ref{eq:covariantconstancy}) along $V$ says that they are
time-independent in the gauge (\ref{gaugefixing}). The spatial components of the Eq.~(\ref{nablaPhi}) give\footnote{We have introduced the spatial connection $\hat{\vec{B}} =
\hat{d}q^X \vec{\omega}_X + {\textstyle\frac{1}{2}} g\hat{A}^{I}\vec{P}_{I}$.}
\begin{equation}
\label{constantJ}
\hat{\mathfrak{D}}_{m}\vec{\Phi}_{np} =  0 \,.
\label{eq:phicovconst}
\end{equation}
We can solve this equation for $\xi^{(-)}$ in an arbitrary frame and $SU(2)$ gauge:
\begin{equation}
\xi^{(-)}{}_{mnp}  =
-\hat{\vec{B}}_{m} \cdot \vec{\Phi}_{np} 
-{\textstyle\frac{1}{4}}\partial_{m}\vec{\Phi}_{nq}\cdot\vec{\Phi}_{qp}\,,
\label{eq:embedding}
\end{equation}
This equation expresses the embedding of the $SU(2)$ connection
$\hat{\vec{B}}$ into the anti-self-dual part of the spin connection of the base manifold. Note that the explicit form of this equation is not affected by the presence of tensor fields, see Ref.~\cite{Bellorin:2007yp}. However, we can not conclude that the class of supersymmetric spatial manifolds is the same with or without tensor fields because we have not analyzed the equation of motion yet.

Alternatively, the above condition can be expressed in terms of curvature tensors. This is achieved by studying the integrability condition of Eq.~(\ref{constantJ}). The results for the anti-selfdual part of the Riemann tensor, the Ricci tensor and the scalar curvature are
\begin{eqnarray}
\hat{R}	^{(-)}{}_{mnkl} & = &
{\textstyle\frac{1}{4}} \hat{\mathfrak{D}}_{m} q^{X} \hat{\mathfrak{D}}_{n} q^{Y} \vec{J}_{XY} \cdot  \vec{\Phi}_{kl}
-{\textstyle\frac{1}{2}} g \hat{F}^{I}{}_{mn} \vec{P}_{I} \cdot  \vec{\Phi}_{kl}
\,,
\label{eq:R-} 
\\ \nonumber \\
\hat{R}_{mn} & = &
-{\textstyle\frac{1}{2}} \hat{\mathfrak{D}}_{m} q^{X}
\hat{\mathfrak{D}}_{n}q^{Y}g_{XY}
+2g^2 f^{-1} C^{\tilde{I}JK} h_{\tilde{I}} \vec{P}_{J}\cdot\vec{P}_{K} \delta_{mn} 
+g \hat{F}^{I(+)}{}_{mp}\vec{\Phi}_{pn}\cdot\vec{P}_{I} \, ,
\label{eq:Ricci} 
\\ \nonumber \\
\hat{R} & = &
-{\textstyle\frac{1}{2}} \hat{\mathfrak{D}}_{m} q^{X}
\hat{\mathfrak{D}}_{m} q^{Y}g_{XY}
+8g^2 f^{-1} C^{\tilde{I}JK} h_{\tilde{I}} \vec{P}_{J}\cdot\vec{P}_{K}\, .
\label{eq:Ricciscalar}
\end{eqnarray}

Previous results on supersymmetric configurations can be deduced as limiting cases of the above relations for the spatial metric $h_{\underline{mn}}$. For example, by sending the coupling constant $g$ to zero we recover the spatial curvature for supersymmetric configurations of the ungauged theory \cite{Bellorin:2006yr}
\begin{equation}
\hat{R}	^{(-)}{}_{mnkl}  = 
{\textstyle\frac{1}{4}} \hat{\partial}_{m} q^{X} \hat{\partial}_{n} q^{Y} \vec{J}_{XY} \cdot \vec{\Phi}_{kl} \,,
\hspace{2em}
\hat{R}_{mn}(h)   = 
-{\textstyle\frac{1}{2}} \hat{\partial}_{m} q^{X}
\hat{\partial}_{n}q^{Y}g_{XY} \,.
\end{equation}
If we also turn off the hyperscalars the resulting spatial manifold is Ricci-flat and Eq.~(\ref{eq:phicovconst}) becomes $\hat{\nabla}_m \vec{\Phi}_{np} = 0$. Therefore $\vec{\Phi}_m{}^n$ is a triplet of integrable complex structures with imaginary unit quaternion algebra. This corresponds to a hyperK\"ahler manifold, as was found in Ref.~\cite{Gauntlett:2002nw}.

Other known limit is the gauged theory without tensor and hypermultiplets. We recall that in this case momentum maps are constant, which take the form $\vec{P}_I =  \vec{e} \, \xi_I$ for the case of the $G$-group that includes $U(1)$ factors. For these models the Eq.~(\ref{eq:phicovconst}) becomes
\begin{equation}
\hat{\nabla}_m \vec{\Phi}_{np} + g \hat{A}^I_ m \xi_I \vec{e} \times \vec{\Phi}_{np}
= 0 \,.
\end{equation}
From this equation it is clear that the tensor $\vec{e}\cdot\vec{\Phi}$ represents an integrable complex structure. Therefore the base spatial manifold is K\"ahler, as was found in Refs.~\cite{Gauntlett:2003fk,Gutowski:2004yv}.


\subsubsection{Solving the Killing spinor equations}

The necessary conditions for having unbroken
supersymmetry that we have derived in the previous section are also
sufficient. Indeed, it can be shown that the $\delta_\epsilon \psi_\mu^i$, $\delta_\epsilon \lambda^{xi}$ and $\delta_\epsilon \zeta^A$ equations are solved by the configurations as we have them, in an arbitrary frame and $SU(2)$ gauge, by the Killing spinor
\begin{equation}
\epsilon^{i}(x,x_0) = 
\sqrt{f} P\exp\left(
-{\textstyle\frac{1}{16}}\int\limits_{x_{0}}^{x} dx_{1}^{\underline{m}}
\partial_{\underline{m}}\!\not\!\Phi_{j}{}^{i}(x_{1})
\right)\epsilon_{0}^{j}\, ,
\end{equation}
where $\epsilon_{0}^{j}$ is a constant spinor. These spinors are subject to the projections
\begin{equation}
\label{eq:projections}
\vec{\Pi}^{+}{}_{j}{}^{i} \epsilon^{j} =0\, ,
\hspace{2em}
R^{-}\epsilon^{i} =0\, ,
\end{equation}
where
\begin{equation}
R^{\pm} \equiv {\textstyle\frac{1}{2}} \left(1\pm\gamma^{0}\right)\, ,
\hspace{1.5cm}
\Pi^{r\pm}{}_{j}{}^{i} \equiv
{\textstyle\frac{1}{2}}\left(\ \delta
\ \pm\ {\textstyle\frac{i}{4}}
\not\!\Phi^{(r)}\sigma^{(r)}\right)_{j}{}^{i}\, .
\end{equation}
In a frame and $SU(2)$ gauge where $\vec{\Phi}$ is constant, the Killing spinor is just $\epsilon^i = \sqrt{f}\epsilon_0^i$.

The supersymmetric configurations preserve in general $1/8$ of the supersymmetries.
These projections are to be imposed for general supersymmetric configurations. There could be, however, special configurations for which some or all of these projections are not needed, hence preserving a big fraction of supersymmetry.


\subsubsection{Supersymmetric solutions}
In the time-like case we find that the independent KSIs (written in the flat five-dimensional frame) are 
\begin{eqnarray}
\mathcal{E}_X 
& = &
0 \,,
\label{eq:ksiq}
\\ \nonumber \\
\mathcal{E}_{ab} 
& = &
-{\textstyle\frac{\sqrt{3}}{2}} \delta_{0(a|} (h^I {\tilde\mathcal{E}}_{I|b)}
    -{\textstyle\frac{1}{2}} h^M \mathfrak{D}^c \mathcal{E}_{Mc|b)}) \,,
\label{eq:ksi8} 
\\ \nonumber \\
\mathcal{E}_x
& = &
h^I_x {\tilde\mathcal{E}}_{I0}
-{\textstyle\frac{1}{2}} h^M_x \mathfrak{D}^a \mathcal{E}_{Ma0} \,,
\label{eq:ksi10}
\\ \nonumber \\
h^I {\tilde\mathcal{E}}_{Im}
& = & 
{\textstyle\frac{1}{2}} h^M \mathfrak{D}^a \mathcal{E}_{Mam}
-2g \Omega^{MN}h_M\mathcal{E}_{N0m} \,,
\label{eq:ksi9}
\\ \nonumber \\
h^I_x {\tilde\mathcal{E}}_{Im}
& = & 
{\textstyle\frac{1}{2}} h^M_x \mathfrak{D}^a \mathcal{E}_{Mam}
+g \Omega^{MN}h_{Mx}\mathcal{E}_{N0m} \,,
\label{eq:ksi10b}
\\ \nonumber \\
g\Omega^{MN} h_M \mathcal{E}_{Nmn} 
& = &
0 \,,
\label{eq:ksi9b} 
\\ \nonumber \\
g\Omega^{MN} h_{Mx} \mathcal{E}_{Nmn} \vec{\Phi}^{mn}
& = &
0 \,.
\label{eq:ksi11}
\end{eqnarray}
Eq.~(\ref{eq:ksiq}) says that all the supersymmetric configurations automatically solve the equation of motion of the hyperscalars. Eqs.~(\ref{eq:ksi8}) and (\ref{eq:ksi10}) imply that if the $\tilde\mathcal{E}_I$ and $\mathcal{E}_M$ equations are satisfied, then the Einstein and $\mathcal{E}_x$ equations are also satisfied. Conditions (\ref{eq:ksi9}) and (\ref{eq:ksi10b}) imply that the space-like components of the equations $\tilde\mathcal{E}_I$ are satisfied if the tensor equation is solved. Finally, Eqs.~(\ref{eq:ksi9b}) and (\ref{eq:ksi11}) say that the anti-self-dual part of the space-like components of the tensor equation, $\mathcal{E}_M^{mn(-)}$, vanishes automatically for supersymmetric configurations. Therefore, in the timelike case, the necessary and sufficient condition for a supersymmetric configuration be also a solution of the theory is that it must solve the time-like component of the vector equation $\tilde\mathcal{E}_I$ and the $\mathcal{E}_M^{0m}$ and $\mathcal{E}_M^{mn(+)}$ components of the tensor equation of motion. 

The time-like components of the Maxwell and $\mathcal{E}_M$ equations evaluated on
supersymmetric configurations yield
\begin{eqnarray}
-{\textstyle\frac{1}{\sqrt3}} f^{-2} \tilde{\mathcal{E}}_I^0 & = &
\hat{\mathfrak{D}}^{2} (h_I/f)
-{\textstyle\frac{1}{12}}C_{I\tilde J\tilde K} \epsilon^{mnpq}
\hat{H}^{\tilde J}_{mn} \hat{H}^{\tilde K}_{pq}
+{\textstyle\frac{1}{\sqrt{3}}} g [ \vec{P}_I \cdot \vec{\Phi}^{mn} (d\omega)^{(-)}_{mn}
+ g f^{-2} k^{\tilde{x}} k_{I\tilde{x}} ] \, , 
\nonumber \\ 
\label{eq:hifequation}
\\ \nonumber \\
f^{-1} \mathcal{E}_M^{0m} v_m & = &
g^{-1} \Omega_{MN} \hat{\star} \hat{\mathfrak{D}} \hat{B}^N
+2\sqrt3 \hat{\mathfrak{D}} (h_M/f)  \,.
\label{eq:dB}
\end{eqnarray}
It can be checked by direct computations that the space-like components of the tensor equation vanish automatically for supersymmetric configurations. From the analysis of the KSIs we were aware of the vanishing of $\mathcal{E}_M^{mn(-)}$.

The Bianchi identity for the field strength $F^I$ holds automatically since we assume that we are always dealing with the gauge potential ${A}^I$ explicitly.

We summarize our results on supersymmetric solutions in the time-like case. The objects that have to be chosen are a scalar function $f$, a 4-dimensional Riemannian manifold with metric $h_{\underline{m}\underline{n}}$ together with an almost quaternion structure $\vec{\Phi}_{mn}$, a 1-form $\omega_{\underline{m}}$, $n_V + n_T$ scalars mappings $\phi^x$ to $\mathcal{M}_{VS}$, $4 n_{H}$ hyperscalar mappings $q^{X}$ to $\mathcal{M}_{QK}$, a gauge potential $\hat{A}^{I}{}_{\underline{m}}$ and $n_T$ two-forms $\hat{B}^M$. All these variables are spatial objects and are independent of time. They have to satisfy the following equations in order to get a configuration with preserved supersymmetry:
\begin{eqnarray}
\xi^{(-)}{}_{mnp}  & = &
-( \hat{\partial}_m q^X \vec{\omega}_X 
  + {\textstyle\frac{1}{2}} g \hat{A}^I_m \vec{P}_I ) \cdot \vec{\Phi}_{np}  \, ,
\label{eq:rspin}
\\ \nonumber \\
%
\hat{\mathfrak{D}}_{m} q^{X} & = &
\Phi^{r}{}_{m}{}^{n} \hat{\mathfrak{D}}_{n} q^{Y} J^{r}{}_{Y}{}^{X}\, ,
\label{eq:rquaternionicmap}
\\ \nonumber \\
%
 h_{\tilde{I}}\hat{H}^{\tilde{I}(+)} & = &  
{\textstyle\frac{2}{\sqrt{3}}} f (d\omega)^{(+)} \, , 
\label{eq:rH+}
\\ \nonumber  \\
\hat{H}^{\tilde{I}(-)} & = & 
-2gf^{-1}C^{\tilde{I}\tilde{J}K}h_{\tilde{J}}\vec{P}_{K}\cdot\vec{\Phi}\, ,
\label{eq:rH-}
\end{eqnarray}
%
where we have formulated these conditions in a frame in which $\vec{\Phi}$ is constant. In addition they have to satisfy the following equation of motion in order to be a solution of the theory:
\begin{eqnarray}
\hat{\mathfrak{D}}^{2} (h_I/f)
-{\textstyle\frac{1}{12}}C_{I\tilde J\tilde K} \epsilon^{mnpq}
\hat{H}^{\tilde J}_{mn} \hat{H}^{\tilde K}_{pq}
+{\textstyle\frac{1}{\sqrt{3}}} g [ \vec{P}_I \cdot \vec{\Phi}^{mn} (d\omega)^{(-)}_{mn}
+ g f^{-2} k^{\tilde{x}} k_{I\tilde{x}} ]
& = & 0 \,,
\label{eq:rmaxwell}
\\ \nonumber \\
\hat{\mathfrak{D}} \hat{B}^M
+2\sqrt3 g \Omega^{MN} \hat{\star} \hat{\mathfrak{D}} (h_N/f) 
& = & 0 \,.
\label{eq:rdb}
\end{eqnarray}
The supersymmetric space-time metric, vector and tensor fields are determined in terms of these variables by
\begin{eqnarray}
 ds^2 & = &
  f^2 (dt +\omega)^2 -f^{-1} h_{\underline{mn}} 
    dx^{\underline{m}} dx^{\underline{n}} \,,
 \label{eq:rmetric}
 \\ \nonumber \\
 A^{I} & = & -\sqrt{3} h^{I} e^0 + \hat{A}^{I} \,,
 \label{eq:rpotential}
 \\ \nonumber \\
 H^{\tilde{I}} & = &  -\sqrt{3} \ 
 \hat{\mathfrak{D}}(h^{\tilde{I}}e^{0})+\hat{H}^{\tilde{I}} \,,
 \label{eq:rFI}
\end{eqnarray}
where $e^0 = f(dt + \omega)$.

\subsection{The null case}
\label{sec-null}

\subsubsection{The equations for the bilinears}
The five-dimensional metric can be put in the form
\begin{equation}
\label{eq:nullmetric}
ds^{2}= 2fdu(dv+Hdu+\omega)
-f^{-2}\gamma_{\underline{r}\underline{s}}dx^{r}dx^{s}\, ,  
\hspace{2em}
l_{\mu}dx^{\mu}= f du\, ,
\hspace{2em}
l^{\mu}\partial_{\mu} =\partial_{\underline{v}}\, ,  
\end{equation}
$l^\mu$ is a null Killing vector, $r,s,t=1,2,3$, denote part of the spatial directions which we call the transverse directions. The objects defining this metric may depend on $u$ but not on $v$. Now with the partial gauge fixing~(\ref{gaugefixing}) we have $A^{I}_{\underline{v}}=0$ and also the scalars field are $v$-independent. By a rotation of the dreibeins preserving the orientation\footnote{We use $\epsilon^{+-123}=\epsilon_{+-123}=+1$. The minus sign in the expression of $\Phi^r$ is needed for the consistency of the equations for the bilinears.} we can bring $\vec{\Phi}$ to the form
\begin{equation}
\Phi^r = - du \wedge v^r \,.
\label{eq:null2-forms}
\end{equation}

The splitting of the gauge potential is
\begin{equation}
A^{I} = A^{I}{}_{\underline{u}} du + \hat{A}^{I} \, ,
\label{eq:potentialdecomposition}
\end{equation}
where $\hat{A}$ is a spatial one-form.  

Eq.~(\ref{eq:covariantconstancy}) becomes
\begin{equation}
du\wedge\left[ dv^{r} 
- \left( 2\varepsilon^{rst} \hat{B}^{t} 
+ 	\sqrt{6} g f^{-1} P^{s} v^{r}\right)\wedge v^{s} \right]
=0\, .
\end{equation}
From this equation we may solve the transverse spin connection
\begin{equation}
\varpi^{rs} =  
2\varepsilon^{rst}\hat{B}^{t}
-2\sqrt{6} g f^{-1} P^{[r} v^{s]}\, .
\label{eq:FixOmegaNull}
\end{equation}

The differential equation for the hyperscalars is\footnote{The covariant derivative
$\hat{\mathfrak{D}}$ includes the transverse gauge connection $\hat{A}^{I}$.}
\begin{equation}
\hat{\mathfrak{D}}_{r} q^{X} J^{r}{}_{X}{}^{Y} = 
- g f^{-1} k^{Y}  \, .
\label{eq:conditiononqx}
\end{equation}

The vector/tensor fields are determined from Eqs.~(\ref{df}), (\ref{dV}), (\ref{nablaPhi}), (\ref{dphi}), (\ref{eq:Phidphi}) and (\ref{FtimesPhi}). Eqs.~(\ref{df}) and (\ref{dphi}) lead to
\begin{equation}
 H^{\tilde I}_{+-}  = 
 g k^x h^{\tilde I}_x \,,
 \hspace{2em}
 H^{\tilde I}_{r-}  = 
 0 \,.
\label{eq:ncHminus}
\end{equation}
The first of these equations is equivalent to
\begin{equation}
F^I_{+-} = 0 \,,
\hspace{2em}
B^M_{+-}
=
2\sqrt{3} g \Omega^{MN} h_N \,,
\end{equation}
where we have made use of the identity (\ref{eq:hk}) and the closure property of $\{h^{\tilde I},h^{\tilde I}_x\}$. The vanishing of the component $F^I_{+-}$, and consequently the vanishing of the component $F^I_{\underline{uv}}$, is consistent with our gauge fixing $A_{\underline{v}}=0$ and the expected $v$-independence of the supersymmetric configurations. Eq.~(\ref{FtimesPhi}) is automatically solved by (\ref{eq:ncHminus}). Therefore, the supersymmetric vector/tensor fields have the general form
\begin{eqnarray}
H^{\tilde I} 
& = & 
H^{\tilde I}_{+-} e^+ e^-  
+ H^{\tilde I}{}_{+r} e^+ e^r 
+ \hat{H}^{\tilde I} \,,
\end{eqnarray}
where $H^{\tilde I}_{+-} = \delta^{\tilde I}_M ( 2\sqrt{3} g \Omega^{MN} h_N)$ and $\hat{H}^{\tilde I}$ are two-forms living in the transverse space,
\begin{equation}
\hat{H}^{\tilde I} 
= 
{\textstyle\frac{1}{2}} \hat{H}^{\tilde I}_{\underline{rs}} 
dx^{\underline r} dx^{\underline s}
\equiv
{\textstyle\frac{1}{2}} \hat{H}^{\tilde I}_{rs} 
v^r v^s \,.
\end{equation}
The transverse components $\hat{H}^{\tilde{I}}$ can be straightforwardly determined following the same steps done in Ref.~\cite{Bellorin:2007yp} to determine $\hat{F}^{I}$. The result is
\begin{equation}
 \hat{H}^{\tilde I}  = 
 \sqrt{3} \ \hat{\star} ( \hat{\mathfrak{D}}K^{\tilde I}
 - {\textstyle\frac{2}{\sqrt{3}}} g f^{-2} \hat{P}^{\tilde I} )\,,
\label{eq:transverseH}
\end{equation}
where $K^{\tilde I} \equiv h^{\tilde I}/f$ and $\hat{P}^{\tilde I} \equiv a^{\tilde{I}J} \hat{P}_J \equiv a^{\tilde{I}J} {P}^r_J v^r$. Similarly to the time-like case, we interpret this equation as a condition for $\hat{A}^I$ rather than $\hat{F}^I$, hence we do not impose the Bianchi identity on the r.h.s. The components $h_{\tilde{I}} H^{\tilde{I}}_{+r}$ are determined from the $++r$ components of Eq.~(\ref{nablaPhi}),
\begin{equation}
h_{\tilde I} H^{\tilde I}{}_{+r} =
-{\textstyle\frac{1}{\sqrt{3}}} f^{2} (\hat{\star} \hat{d}\omega)_{r}
\, ,
\label{eq:FIprNull}
\end{equation}
where we have assumed that
\begin{equation}
 B^r_{\underline{u}} 
 +{\textstyle\frac{1}{4}} \epsilon_{rst} v_s{}^{\underline s}
     \partial_{\underline u} v_{t\underline s}  
= 0 \,,
\label{eq:NullAConsistency}
\end{equation}
which is nothing but a partial gauge fixing condition that affects the freedom to rotate the dreibeins. This condition basically fixes the $u$-dependence of the dreibeins. We can solve the one-form $\omega$ in terms of the vector and tensor fields from the equation (\ref{eq:FIprNull}),
\begin{equation}
\hat{d}\omega =
\sqrt{3} f^{-2} \ \hat{\star} (
h_{I} \hat{\mathfrak{D}} A^{I}{}_{\underline{u}} 
-h_M \hat{b}^M 
- h_{I} \partial_{\underline{u}}\hat{A}^{I} ) \,,
\label{eq:omega}
\end{equation}
where we have defined the transverse one-forms $\hat{b}^M \equiv B^M{}_{+r} v^r$. This equation fixes $\omega$ up to an arbitrary gradient. Therefore, the supersymmetric vector/tensor fields must take the form
\begin{eqnarray}
F^{I} 
&=& 
( \hat\mathfrak{D} A^I_{\underline{u}} 
  - \partial_{\underline{u}} \hat{A}^I ) \wedge du
+ \hat{F}^I \,,
\label{eq:vectorfieldstrengths}
\\ \nonumber \\
B^M
& = &
2\sqrt{3} g f \Omega^{MN} h_N du\wedge (dv+\omega)
-\hat{b}^M \wedge du
+ \hat{B}^M \,,
\label{eq:nctensors}
\end{eqnarray}
where $\hat{F}^I$ and $\hat{B}^M$ are given in Eq.~(\ref{eq:transverseH}).


\subsubsection{Solving the Killing spinor equations}
\label{sec:SolvKSENull}
After use of all the information extracted from the equations for the bilinears, we conclude that the KSEs are solved by constant spinors subject to the projections
\begin{equation}
\gamma^+ \epsilon^i = 0 \,,
\hspace*{2em}
\Pi^{r(+)}{}_i{}^j \epsilon^i = 0 \,,
\end{equation}
where
\begin{equation}
\Pi^{r(\pm)}{}_i{}^j 
\equiv
{\textstyle\frac{1}{2}} (\delta_i{}^j \pm i\gamma^{(r)}\sigma^{(r)}{}_i{}^j ) \,.
\end{equation}
These configurations preserve $1/8$ of the supersymmetries.

\subsubsection{Equations of motion}
In the null case the independent KSIs are
\begin{eqnarray}
\mathcal{E}_X 
& = & 0 \,,
\label{eq:ksinull1}
\hspace{55mm}
\mathcal{E}_{rs} 
\ = \ 
0\,,
\\ \nonumber \\
\mathcal{E}_x 
& = & 
g \Omega^{MN} h_{Mx} \mathcal{E}_{N+-} \,,
\label{eq:ksinull2}
\\ \nonumber \\
\mathcal{E}_{r+} 
& = & 
{\textstyle\frac{\sqrt3}{2}} g \Omega^{MN} h_M \mathcal{E}_{Nr+} \,,
\label{eq:ksinull4}
\hspace{25mm}
\mathcal{E}_{a-} 
\ = \ 
{\textstyle\frac{\sqrt3}{2}} g \Omega^{MN} h_M \mathcal{E}_{Na-} \,,
\\ \nonumber \\
h^I {\tilde\mathcal{E}}_{Ia}
& = & 
{\textstyle\frac{1}{2}} h^M \mathfrak{D}^b \mathcal{E}_{Mba} \,,
\label{eq:ksinull5}
\hspace{30mm}
h^I_x {\tilde\mathcal{E}}_{I-}
\ = \ 
{\textstyle\frac{1}{2}} h^M_x \mathfrak{D}^b \mathcal{E}_{Mb-} \,,
\\ \nonumber \\
h^I_x {\tilde\mathcal{E}}_{Ir}
& = & 
{\textstyle\frac{1}{2}} h^M_x \mathfrak{D}^b \mathcal{E}_{Mbr} 
+{\textstyle\frac{1}{2}} g \Omega^{MN} h_{Mx} \mathcal{E}_{Nst} \epsilon_{rst}\,,
\label{eq:ksinull7}
\\ \nonumber \\
g \Omega^{MN} h_M \mathcal{E}_{Nrs} 
& = & 
0\,,
\label{eq:ksinull10}
\hspace{32mm}
g\Omega^{MN} h_{Mx} \mathcal{E}_{Nr-} 
\ = \ 
0\,.
\end{eqnarray}
Let us analyze this system. $\mathcal{E}_X$ is automatically on-shell for supersymmetric configurations, as well as the transverse components $\mathcal{E}_{rs}$ of the Einstein equations. If the tensor equation of motion is satisfied, Eqs.~(\ref{eq:ksinull2}) and (\ref{eq:ksinull4}) imply that the $\mathcal{E}_x$, $\mathcal{E}_{a-}$ and $\mathcal{E}_{r+}$ equations are also satisfied, whereas Eqs.~(\ref{eq:ksinull5})-(\ref{eq:ksinull7}) imply that the components ${\tilde\mathcal{E}}_{I-}$ and ${\tilde\mathcal{E}}_{Ir}$ of the Maxwell equations are satisfied. Finally, Eqs.~(\ref{eq:ksinull10}) say that some projections of the tensor equation of motion are automatically on-shell for supersymmetric configurations. Therefore, in the null case, the independent equations of motion for supersymmetric configurations are $\mathcal{E}_{++}$, ${\tilde\mathcal{E}}_{I+}$ and the tensor equation of motion, $\mathcal{E}_M$.

The $\star\mathcal{E}_{M\underline{rst}}$ and $\star\mathcal{E}_{M\underline{urs}}$ components yield respectively
\begin{eqnarray}
\hat{\mathfrak{D}}\hat{\star}\hat{\mathfrak{D}} K^M
-{\textstyle\frac{2}{\sqrt{3}}} g 
\hat{\mathfrak{D}} \hat{\star} (f^{-2} \hat{P}^M)
\ = \ 
-4\sqrt{3} g^2 \Omega^{MN} M_N \mathrm{vol_3} &\,,& 
\label{eq:nobianchi}
\\ \nonumber \\
\hat{\mathfrak{D}} \hat{b}^M 
- \mathfrak{D}_{\underline{u}} \hat{B}^M
+\sqrt3 g K^{\tilde I} \hat{\star} \left[
  2 t_{(\tilde I J)}{}^M ( \partial_{\underline{u}} \hat{A}^J
    -\hat{\mathfrak{D}} A^J{}_{\underline{u}} )
    + t_{\tilde I N}{}^M  \hat{b}^N \right]
 \ = \  0 &\,,&
\label{eq:db}
\end{eqnarray}
where
\begin{equation}
M_{\tilde I}  \equiv 
f^{-3} t_{(\tilde I \tilde J)}{}^M h_M h^{\tilde J} \,,
\end{equation}
and $\mathrm{vol_3}$ is the volume element of the three-dimensional transverse space. The rest of the components of the $\mathcal{E}_M$ vanish automatically.

The $\tilde{\mathcal{E}}_{I+}$ component of the Maxwell equations (\ref{eq:ERm}) yields
\begin{equation}
\begin{array}{l}
f^{-2} \mathcal{E}_{I+} \mathrm{vol_3} 
\ = \   
2 C_{I\tilde{J}K} \left[
   \hat{\mathfrak{D}}\hat{\star}
      (K^{\tilde{J}}\hat{\mathfrak{D}}A^K{}_{\underline{u}})         
  -\hat{\mathfrak{D}} \hat{\star} 
    (K^{\tilde J} \partial_{\underline{u}}\hat{A}^K)    
     +{\textstyle\frac{1}{\sqrt{3}}} \hat{H}^{\tilde{J}} \wedge
     ( \hat{\mathfrak{D}} A^K{}_{\underline{u}} 
         - \partial_{\underline{u}} \hat{A}^K ) \right]
\\ \\
-2 C_{I\tilde J M} \left[
  \hat{\mathfrak{D}} \hat{\star} (K^{\tilde J} \hat{b}^M)  
    + {\textstyle\frac{1}{\sqrt{3}}} \hat{H}^{\tilde{J}}
        \wedge \hat{b}^M  \right]
+2 g \hat{P}_{I}\wedge \hat{d}\omega
+ g f^{-3}k_{I\,\tilde x} \mathfrak{D}_{\underline{u}}\varphi^{\tilde x} 
\mathrm{vol_3} 
+6 g \mathfrak{D}_{\underline{u}} \hat{\star} M_I\,.
\end{array}
\label{eq:ncgensusymaxwell}
\end{equation}
This equation is $G$-symmetry invariant, in particular, under $u$-dependent transformations. This fact can be used to partially fix the $G$-gauge by imposing
\begin{equation}
C_{I\tilde{J}K} \left[ 
 \hat{\mathfrak{D}} \hat{\star} 
    (K^{\tilde J} \partial_{\underline{u}}\hat{A}^K)
  + {\textstyle\frac{1}{\sqrt{3}}} \hat{H}^{\tilde{J}}
        \wedge \partial_{\underline{u}} \hat{A}^K \right]
-{\textstyle\frac{1}{2}} g f^{-3} k_{I\tilde{x}} \dot{\varphi}^{\tilde{x}} \mathrm{vol_3}
-3 g \partial_{\underline{u}} \hat{\star} M_I
= 0\,,
\label{eq:Ggauge}
\end{equation}
such that the Eq.~(\ref{eq:ncgensusymaxwell}) becomes
\begin{equation}
\begin{array}{rcl}
f^{-2} \mathcal{E}_{I+} \mathrm{vol_3} 
& = &   
2 C_{I\tilde{J}K} \left[
 \hat{\mathfrak{D}}\hat{\star}
 (K^{\tilde{J}}\hat{\mathfrak{D}} A^K{}_{\underline{u}})
   +{\textstyle\frac{1}{\sqrt{3}}} \hat{H}^{\tilde{J}} \wedge 
     \hat{\mathfrak{D}} A^K{}_{\underline{u}} \right]
 -2 C_{I\tilde{J}M} \left[
     \hat{\mathfrak{D}} \hat{\star} (K^{\tilde J} \hat{b}^M)
    +{\textstyle\frac{1}{\sqrt{3}}} \hat{H}^{\tilde{J}}
        \wedge \hat{b}^M  \right]   
\\ \\
&   & +2 g \hat{P}_{I}\wedge \hat{d}\omega
+ g^2 A^J{}_{\underline{u}}( 
 f^{-3} k_J{}^{\tilde{x}} k_{I\,\tilde x}
  -6 t_{JI}{}^{\tilde{K}} M_{\tilde K} ) \mathrm{vol_3} \,,
\end{array}
\label{eq:ncsusymaxwell}
\end{equation}
which is still covariant under $u$-independent $G$-symmetry transformations. In absence of tensor fields the term
\begin{equation}
C_{IJK}\left[
\hat{\mathfrak{D}}\hat{\star}
      (K^{J}\hat{\mathfrak{D}}A^K{}_{\underline{u}})
+\hat{\mathfrak{D}} K^{J} \wedge \hat{\star} 
  \hat{\mathfrak{D}}A^K{}_{\underline{u}}
    \right]
\end{equation}
can be expressed in terms of $G$-covariant Laplacians of $K^I$ and $C_{IJK} K^J A^K_{\underline{u}}$. In presence of tensor fields, however, this term is no longer covariant, hence we cannot write it in terms of covariant Laplacians. This obstruction is a consequence of the fact that tensor fields, which are the $G$-symmetry partners of vector fields, are not derived from any gauge potentials.

Now we turn our attention to the $\mathcal{E}_{++}$ component of the Einstein equation. There is a gauge freedom in the expression (\ref{eq:omega}) for the one-form $\omega$, which we use to impose a condition on $\omega$:
\begin{equation}
\begin{array}{l}
\hspace*{-10mm}
\nabla_{r}(\dot\omega)_{r} + 3(\dot\omega)_{r}\partial_{r} \log f  = 
\\ \\
- {\textstyle\frac{1}{2}}f^{-3}(\ddot\gamma)_{rr}
- {\textstyle\frac{1}{4}}f^{-3}(\dot\gamma)^2\nonumber
+ {\textstyle\frac{3}{2}} f^{-4}\dot f(\dot\gamma)_{rr}
+ 3f^{-3}[\partial_{\underline{u}}^2 \log f -2(\partial_{\underline{u}}\log f)^2]
\\ \\
-{\textstyle\frac{1}{2}}f^{-3}\left[ 
 g_{\tilde{x}\tilde{y}} (\dot \varphi^{\tilde{x}} \dot \varphi^{\tilde{y}}  
+2 g \dot{\varphi}^{\tilde{x}} A^{I}{}_{\underline{u}} k_{I}{}^{\tilde{y}} )
\right]
\\ \\
+C_{\tilde{I}JK} K^{\tilde{I}} \left[
 (\partial_{\underline{u}}\hat{A}^{J})_{r} 
(\partial_{\underline{u}}\hat{A}^{K})_{r}
-2 \hat{\mathfrak{D}}_{r} A^{J}{}_{\underline{u}} 
(\partial_{\underline{u}}\hat{A}^{K})_{r} \right]
+2 C_{\tilde{I}JM} K^{\tilde{I}} (\partial_{\underline{u}} \hat{A}^J)_r
  \hat{b}^M{}_r
\, .  
\end{array}
\label{eq:fixomega} 
\end{equation}
The $\mathcal{E}_{++}$ component of the Einstein equations becomes
\begin{equation}
\begin{array}{rcl}
-f^{-1}\mathcal{E}_{++} 
& = &
\hat{\nabla}^2 H 
+ C_{\tilde{I}JK} K^{\tilde{I}} \hat{\mathfrak{D}}_r A^J{}_{\underline{u}}
   \hat{\mathfrak{D}}_r A^K{}_{\underline{u}}
-2 C_{\tilde{I}JM} K^{\tilde{I}} \hat{b}^M{}_r 
   \hat{\mathfrak{D}}_r A^J{}_{\underline{u}}
+C_{\tilde{I}MN} \hat{b}^M{}_r \hat{b}^N{}_r
\\ \\ &  &
-{\textstyle\frac{1}{2}} g^2 f^{-3}A^{I}{}_{\underline{u}}
  A^{J}{}_{\underline{u}} g_{\tilde{x}\tilde{y}} k_{I}{}^{\tilde{x}} 
    k_{J}{}^{\tilde{y}}
\end{array}
\label{eq:einstein++}
\end{equation}

Let us summarize the results of the null case. The objects that have to be chosen are two functions $f$ and $H$, a 3-dimensional transverse metric $\gamma_{\underline{r}\underline{s}}$ together with a Driebein basis $v^{r}$ for it, $n_V + n_T$ scalar mappings $\phi^x$ to $\mathcal{M}_{VS}$, $4 n_{H}$ hyperscalar mappings $q^{X}$ to $\mathcal{M}_{QK}$, the components $\hat{A}^{I}$ and $A^{I}{}_{\underline{u}}$ of the gauge connection $A^{I}$ and $n_T$ transverse one-forms $\hat{b}^M$. All these variables may depend on $u$ but must be $v$-independent. They must satisfy the following equations for preserved supersymmetry:
\begin{eqnarray}
		\varpi^{rs} & = & 
		2 \varepsilon^{rst} (\hat{d}q^X \omega_X^t +
		    {\textstyle\frac{1}{2}} g \hat{A}^I P_I^t  ) 
		       - 2\sqrt{6} g f^{-1} P^{[r} v^{s]}\, ,		
		\label{eq:rnullspin}
		\\ \nonumber \\
%
	\hat{\mathfrak{D}}_{r} q^{X} J^{r}{}_{X}{}^{Y} & = & 
		- g f^{-1} k^{Y}  \, ,
\label{eq:rnullquaternionic}
	\\ \nonumber \\
		\hat{F}^{I}  & = & 
 			\sqrt{3} \ \hat{\star} ( \hat{\mathfrak{D}}K^{I}
 			- {\textstyle\frac{2}{\sqrt{3}}} g f^{-2} \hat{P}^{I} )\,,
 \label{eq:rF}
\end{eqnarray}
where $K^{\tilde{I}} = h^{\tilde{I}}/f$. The equations of motion to be imposed are
\begin{eqnarray}
		\hat{\mathfrak{D}}\hat{\star}\hat{\mathfrak{D}} K^{M} 
		- {\textstyle\frac{2}{\sqrt{3}}}g 
		\hat{\mathfrak{D}}(\hat{\star}f^{-2} \hat{P}^{M})
		 \ = \
		-4\sqrt3 g^2 \Omega^{MN} M_N \mathrm{vol_3} \,, & &
	\label{eq:rtensoreom1}
  \\ \nonumber \\
		\hat{\mathfrak{D}} \hat{b}^M 
			- \mathfrak{D}_{\underline{u}} \hat{B}^M
		+\sqrt3  g K^{\tilde I} \hat{\star} \left[
  	2 t_{(\tilde I J)}{}^M ( \partial_{\underline{u}} \hat{A}^J
    -\hat{\mathfrak{D}} A^J{}_{\underline{u}} )
    + t_{\tilde I N}{}^M  \hat{b}^N \right]
				 \ = \ 0  \,, & &
	\label{eq:rtensoreom2}
	\\ \nonumber \\
		C_{I\tilde{J}K} \left[
 		\hat{\mathfrak{D}}\hat{\star}
 		(K^{\tilde{J}}\hat{\mathfrak{D}} A^K{}_{\underline{u}})
   	+{\textstyle\frac{1}{\sqrt{3}}} \hat{H}^{\tilde{J}} \wedge 
     \hat{\mathfrak{D}} A^K{}_{\underline{u}} \right]
 		- C_{I\tilde{J}M} \left[
     \hat{\mathfrak{D}} \hat{\star} (K^{\tilde J} \hat{b}^M)
    +{\textstyle\frac{1}{\sqrt{3}}} \hat{H}^{\tilde{J}}
        \wedge \hat{b}^M  \right]   & &  \nonumber
\\ \nonumber \\
		+ g \hat{P}_{I}\wedge \hat{d}\omega
		+ {\textstyle\frac{1}{2}} g^2 A^J{}_{\underline{u}}( 
 		f^{-3} k_J{}^{\tilde{x}} k_{I\,\tilde x}
  	-6 t_{JI}{}^{\tilde{K}} M_{\tilde K} ) \mathrm{vol_3} & = & 0 \,, 
  	\nonumber \\
 \label{eq:rmaxwelleom}
\\ \nonumber \\ 
\hat{\nabla}^2 H 
+ C_{\tilde{I}JK} K^{\tilde{I}} \hat{\mathfrak{D}}_r A^J{}_{\underline{u}}
   \hat{\mathfrak{D}}_r A^K{}_{\underline{u}}
-2 C_{\tilde{I}JM} K^{\tilde{I}} \hat{b}^M{}_r 
   \hat{\mathfrak{D}}_r A^J{}_{\underline{u}}
+C_{\tilde{I}MN} \hat{b}^M{}_r \hat{b}^N{}_r &  &  \nonumber 
\\ \nonumber \\ 
-{\textstyle\frac{1}{2}} g^2 f^{-3}A^{I}{}_{\underline{u}}
  A^{J}{}_{\underline{u}} g_{\tilde{x}\tilde{y}} k_{I}{}^{\tilde{x}} 
    k_{J}{}^{\tilde{y}}
& = & 0 \,. \nonumber \\
\label{eq:reinsteineom}
\end{eqnarray}
where $M_{\tilde I} = f^{-3} t_{(\tilde I \tilde J)}{}^M h_M h^{\tilde J}$ and $\hat{B}^M$ and $\hat{d}\omega$ are going to be indicated below. In addition, for $u$-dependent configurations the partial gauge fixings (\ref{eq:NullAConsistency}), (\ref{eq:Ggauge}) and (\ref{eq:fixomega}) must be imposed. The supersymmetric space-time metric, vector and tensor fields are determined in terms of these variables by
\begin{eqnarray}
 ds^{2} & = &
   2 f du (dv + H du + \omega)   
     - f^{-2} \gamma_{\underline{r}\underline{s}} dx^{r} dx^{s}\, ,
 \label{eq:rnullmetric}
\\ \nonumber \\   
 \hat{d}\omega & = &
   \sqrt{3} f^{-2} \ \hat{\star} (
     h_{I} \hat{\mathfrak{D}} A^{I}{}_{\underline{u}} - h_M \hat{b}^M 
       - h_{I} \partial_{\underline{u}}\hat{A}^{I} ) \,,
 \label{eq:rnulldomega}
\\ \nonumber \\
 A^{I} & = & A^{I}{}_{\underline{u}} du + \hat{A}^{I} \, ,
\\ \nonumber \\
 F^{I} & = & 
  ( \hat\mathfrak{D} A^I_{\underline{u}} - \partial_{\underline{u}} \hat{A}^I     ) \wedge du + \hat{F}^I \,,
\\ \nonumber \\
 B^M & = &
   2\sqrt{3} g f \Omega^{MN} h_N du\wedge (dv+\omega)
     -\hat{b}^M \wedge du + \hat{B}^M \,,
 \label{eq:rB}
\\ \nonumber \\
	\hat{B}^{M} & = & 
 			\sqrt{3} \ \hat{\star} ( \hat{\mathfrak{D}}K^{M}
 			- {\textstyle\frac{2}{\sqrt{3}}} g f^{-2} \hat{P}^{M} )\,.
 \label{eq:rhatB}
\end{eqnarray}

\section{Scalar-gravity solutions}
\label{sec:vacuum}
Having performed the characterization of general supersymmetric solutions, we analyze now these conditions for the case of solutions with vanishing vector and tensor fields and constant v.e.v for the hyperscalars. Thus these are configurations with only the scalars $\phi^x$ and the metric $g_{\mu\nu}$ as the non-trivial fields. This study is illustrative since for this kind of configurations several of the conditions for supersymmetric solutions can be solved in a closed way. Moreover, one can go to the limit of pure gravity, which may facilitate future studies of vacuum solution with sources. Also the analysis could be very useful to define appropriated frameworks to search asymptotically vacuum solutions, as domain walls. A similar analysis has been done in Ref.~\cite{Gutowski:2005id} for the null case of the theory without tensor- and hypermultiplets.

\subsection{Time-like case}

We consider the vanishing of vector and tensor fields and also $q^X = \mbox{constant}$. The objects belonging to $\mathcal{M}_{QK}$, such as $g_{XY}$, $k_I^{X}$ and $\vec{P}_I$, are constants subject to the several constraints of the theory.  

In the time-like case it is not convenient to put directly $A^I = 0$ but instead to work in a different gauge which also leads to $F^I =0$. Indeed,
if $H^{\tilde{I}} = 0$ the Eq.~(\ref{eq:rFI}) implies
\begin{eqnarray}
 & & \hat{\mathfrak{D}} ( f h^{\tilde{I}} )  \ = \ 0  
 \hspace{2em} \Rightarrow \hspace{2em}
 \hat{\mathfrak{D}} ( h_{\tilde{I}} / f) \ = \ 0  \,,
\label{eq:constantfh}
\\ \nonumber \\
 & & \hat{H}^{\tilde{I}} = 
  \sqrt{3} f h^{\tilde{I}} d\omega \,.
\label{eq:vacuumH}
\end{eqnarray}
In view of the expression for $\hat{F}^{\tilde{I}}$ given in Eqs.~(\ref{eq:vacuumH}) we choose the gauge
\begin{equation}
\hat{A}^I = \sqrt{3} f h^I \omega \,.
\label{eq:vacuumgauge}
\end{equation}
In this gauge the first of Eqs.~(\ref{eq:constantfh}) becomes
\begin{equation}
  d(f h^{\tilde{I}})  =  
   2 g \delta^{\tilde{I}}{}_M \Omega^{MN} h_N f^2 \omega \,,
 \label{eq:dfhm}
\end{equation}
and one effectively recovers the expression for $\hat{F}^I$ given in Eq.~(\ref{eq:vacuumH}). By projecting the equation (\ref{eq:dfhm}) to $h_{\tilde{I}}$ and $h_{\tilde{I}}^x$ we convert it into two equations
\begin{eqnarray}
 d f & = & 0 \,,
 \\ \nonumber \\
 d \phi^x & = & - g f k^x \omega \,,
 \label{eq:dphi}
\end{eqnarray}
where we have made use of the identity (\ref{eq:hk}). From now on we put $f = 1$ for simplicity. Notice that Eq.~(\ref{eq:dfhm}) implies that the components $h^I$ are constant, thus in the gauge (\ref{eq:vacuumgauge}) the 5d potentials are constant (unphysical) electrostatic potentials, $A^I = - \sqrt{3} h^I dt$. 


If we contrast the expression (\ref{eq:vacuumH}) for $\hat{H}^{\tilde{I}}$ with Eq.~(\ref{eq:rH+}) we obtain that $\hat{H}^{\tilde{I}(+)} = (d\omega)^{(+)} = 0$. If we now compare with Eq.~(\ref{eq:rH-}) we obtain
\begin{equation}
h^{\tilde{I}} d\omega =
-{\textstyle\frac{2}{\sqrt{3}}} g C^{\tilde{I}\tilde{J} K} h_{\tilde{J}}
    \vec{P}_K \cdot \vec{\Phi} \,,
\end{equation}
which can be splitted by projecting out to $h_{\tilde{I}}$ and $h_{\tilde{I}x}$. This yields
\begin{eqnarray}
d\omega  & = & 
-{\textstyle\frac{2\sqrt{2}}{\sqrt{3}}} g \vec{P} \cdot \vec{\Phi} \,,
\label{eq:vacuumdomega}
\\ \nonumber \\
\vec{P}_x  & = & 0 \,.
\end{eqnarray}
We see that in the time-like case there are not supersymmetric scalar-gravity solutions with non-vanishing $\vec{P}_x$ contribution to the potential. We also remark that $\vec{P} \sim h^I \vec{P}_I$ is constant for this kind of configurations.

Now we examine the Maxwell equation (\ref{eq:rmaxwell}). The first term vanish due to Eq.~(\ref{eq:constantfh}). Using the expressions we have obtained for $\hat{H}^{\tilde{I}}$ and $d\omega$ and the fact that $\vec{P}_x = 0$, it is straightforward to see that the second and third terms of the Maxwell equation (\ref{eq:rmaxwell}) cancel out mutually. Thus the Maxwell equation yields (gauged case) $k^{\tilde{x}} k_{I\tilde{x}} = 0 $. Projecting this equation to $h^I$ we obtain
\begin{equation}
g_{\tilde{x}\tilde{y}} k^{\tilde{x}} k^{\tilde{y}} = 0 \,.
\end{equation}
Since it is assumed that the metric of $\mathcal{M}_{VS}$ is Riemannian, we conclude that $k^{\tilde{x}} = 0$. Putting this result back into Eq.~(\ref{eq:dphi}) we obtain that the supersymmetric scalar fields $\phi^x$ are necessarily constant. Consequently, all objects belonging to the target $\mathcal{M}_{VS}$ and its ambient manifold like $h^I$, $h^I_x$ and $g_{xy}$ become constant. Since we have ended up with constant v.e.v. for all the scalar fields and vector/tensor fields vanish, this class of solutions must be critical point of the scalar potential $\mathcal{V}$, as was mentioned at the end of section \ref{sec:theory}. From the point of view of the Einstein equations, these configurations are solutions of the vacuum equations (vanishing energy-momentum tensor) with or without cosmological constant.

It is easy to see that the tensor Eq.~(\ref{eq:rdb}) and the Eq.~(\ref{eq:rquaternionicmap}) for the hyperscalars are automatically solved by the scalar-gravity configurations as we have them.


The form of the five-dimensional metric depends on the triviality of $\omega$, which in turn depends on the vanishing or not of the constant $\vec{P}$. Hence we consider the two cases separately.


\paragraph{Case 1: $\vec{P} = 0$ \\}
In this case Eq.~(\ref{eq:vacuumdomega}) says that $\omega$ is a closed one-form, hence it can be removed from the 5d metric by a coordinate transformation. Thus we set $\omega = 0$ for this class of configurations. According to Eqs.~(\ref{eq:vacuumH}) and (\ref{eq:vacuumgauge}), $\hat{A}^I$ and $\hat{B}^M$ vanish. 



The last condition for preserved supersymmetry is the Eq.~(\ref{eq:rspin}) for the spatial spin connection. In the Case 1 this equation yields that the spin connection is self-dual, $\xi^{(-)}_{mn} = 0$, which correspond to a spatial hyperK\"ahler manifold ($SU(2)$ holonomy). The three complex structures are precisely given by $\vec{\Phi}$, Eq.~(\ref{eq:phicovconst}) establishing their integrability, $\nabla_m \vec{\Phi}_{np} = 0$.

In summary, the configurations in this case are
\begin{equation}
ds^2 = dt^2 - h_{\underline{mn}} dx^{\underline{m}} dx^{\underline{n}} \,,
\hspace{2em}
\phi^x = \mbox{constant}
\end{equation}
where $h_{\underline{mn}}$ is an arbitrary four-dimensional hyperk\"ahler metric. The v.e.v. for the scalars and the hyperscalars are restricted by the conditions $\vec{P}_I = k^{\tilde{x}} = 0$. The simplest five-dimensional solution of this kind, given by $h_{\underline{mn}} = \delta_{\underline{mn}}$, is Minkowski space-time.

The vanishing of $\vec{P}$, $\vec{P}_x$ and $k^{\tilde{x}}$ implies that these configurations have a zero scalar potential. Since the scalar potential is quadratic in these quantities, obviously this class of solutions are critical points of it.


\paragraph{Case 2: $\vec{P} \neq 0$ \\}
Now in this class of solutions the one-form $\omega$ given in Eq.~(\ref{eq:vacuumdomega}) is non-trivial, and also the variables $\hat{A}^I$ do not vanish, as well as the two-forms $\hat{F}^I$ and $\hat{B}^M$. 

From Eq.~(\ref{eq:vacuumdomega}) we see that the non-vanishing two-form $\Omega \equiv \vec{e}_P \cdot \vec{\Phi}$, where $\vec{e}_P$ is the unit vector along $\vec{P}$, is a closed form. From the properties of $\vec{\Phi}$ it is clear that $\vec{e}_P \cdot \vec{\Phi}_m{}^n$ represents an integrable complex structure, thus we have that $h_{\underline{mn}}$ is a K\"ahler metric. This K\"ahler metric is not arbitrary, it must satisfy the condition (\ref{eq:rspin}) on its spin connection, which takes the form
\begin{equation}
\xi^{(-)}_{mnp}  = 
-{\textstyle\frac{\sqrt{3}}{\sqrt{2}}} g P \omega_m \Omega_{np}
\label{eq:vacuumspin}
\,.
\end{equation}
The five-dimensional supersymmetric metric is
\begin{equation}
 ds^2 = (dt+\omega)^2 - h_{\underline{mn}} dx^{\underline{m}}     
        dx^{\underline{n}} \,,
\end{equation}
where $\omega$ and the K\"ahler metric are subject to (\ref{eq:vacuumdomega}) and (\ref{eq:vacuumspin}). The former states that $\omega$ is a local potential for the K\"ahler form. In this case the v.e.v. for the scalars are restricted by the conditions $\vec{P}_I \neq 0$, $\vec{P}_x = k^{\tilde{x}} = 0$.

For this class of configurations the scalar potential takes the value
\begin{equation}
\mathcal{V} = 8 g P^2 > 0 \,.
\end{equation}
To determine the value of its derivatives, take into account that, first, the terms proportional to $P_{x}^2$ and $k^2$ yields a vanishing first derivative once evaluated on the configurations. The derivatives of the $P^2$ term w.r.t. $\phi^x$ also vanish trivially. Finally, taking derivatives of the $P^2$ term w.r.t. $q^X$ we obtain
\begin{equation}
 \partial_X P^2 = 
   {\textstyle\frac{1}{\sqrt{2}}} h^I \vec{P} \cdot \partial_X \vec{P}_I =
     {\textstyle\frac{1}{\sqrt{2}}} h^I \vec{P} \cdot ( 
        -{\textstyle\frac{1}{2}} \vec{J}_{XY} k_I^Y
          -2 \vec{\omega}_X \times \vec{P}_I) = 0 \,,
\end{equation}
where we have used the standard formula for the derivative of a momentum map and the last equality follows after evaluating on the scalar-gravity configurations. As we expected, this class of supersymmetric solutions are critical points of the potential and this is positive (negative cosmological constant).


\subsection{Null case}
Again we consider constant values for the hyperscalars and the vanishing of vector and tensor fields, but now we work directly in the gauge $A^I = 0$, thus we set
\begin{equation}
 \hat{A}^I = A^I_{\underline{u}} = 0 \,.
\end{equation}
We also assume that the configurations are $u$-independent.

Since $F^I$ and $B^M$ are vanishing, we deduce from Eqs.~(\ref{eq:rF}), (\ref{eq:rB}) and (\ref{eq:rhatB}) that $\hat{b}^M  =  0 $, $ h_M  =  0$ and\
\begin{equation}
  \hat{d} ( h^{\tilde{I}} / f )
    - {\textstyle\frac{2}{\sqrt{3}}} g f^{-2} \hat{P}^{\tilde{I}}  =  0 \,,
\end{equation}
which is equivalent to
\begin{equation}
 \hat{d} ( f h_I ) + {\textstyle\frac{2}{\sqrt{3}}} g \hat{P}_I = 0 \,.
 \label{eq:dfhi}
\end{equation}
Projecting this equation to $h^{\tilde{I}}$ yields
\begin{equation}
 \hat{d} f 
  + {\textstyle\frac{2\sqrt{2}}{\sqrt{3}}} g P^r v^r  =  0  \,.
  \label{eq:vacuumdf}
\end{equation}

The condition (\ref{eq:rnullspin}) for the spin connection becomes
\begin{equation}
 \varpi^{rs} = - 2\sqrt{6} g f^{-1} P^{[r} v^{s]} \,.
\end{equation}
By combining this equation with the Cartan's structure equation and Eq.~(\ref{eq:vacuumdf}) it can be shown that 
\begin{equation}
 \hat{d} ( f^{-3/2} v^r ) = 0 \,.
\end{equation}
Thus the transverse one forms $f^{-3/2} v^r$ are locally exact one-forms, $f^{-3/2} v^r = \hat{d} y^r$, for some functions $y^r$. If we take $y^{\underline r} = \delta^{\underline{r}}{}_r y^r$ as the transverse coordinates, then the transverse metric acquires a diagonal form,
\begin{equation}
 \gamma_{\underline{rs}} = f^3 \delta_{\underline{rs}} \,,
\end{equation}
and the Eq.~(\ref{eq:dfhi}) takes the form
\begin{equation}
 f^{-3/2} \partial_{\underline{r}} ( f h_I ) =
   - {\textstyle\frac{2}{\sqrt{3}}} g P_I^{\underline{r}} \,.
\label{eq:rdfhi}
\end{equation}

The last condition for preserved supersymmetry is the Eq.~(\ref{eq:rnullquaternionic}), from it we obtain $k^X = 0$. From the indentity (\ref{eq:hk}) we also see that $h_M = 0$ implies $k^x = 0$, thus again we have the vanishing of $k^{\tilde{x}}$. 

Putting these results in Eq.~(\ref{eq:rnulldomega}) we obtain $\hat{d}\omega = 0$, which, together with the $u$-independence, implies that the one-form $\omega$ can be removed from the metric (\ref{eq:rnullmetric}) by a coordinate transformation.

Now we analyze the supersymmetric equations of motion. It is easy to see that  any scalar-gravity configuration that satisfies the conditions we have found automatically solves the supersymmetric equations of motion (\ref{eq:rtensoreom1}), (\ref{eq:rtensoreom2}) and (\ref{eq:rmaxwelleom}). The Einstein Eq.~(\ref{eq:reinsteineom}) becomes into a harmonicity condition for the scalar $H$,
\begin{equation}
 \hat{\nabla}^2 H = 0 \,.
\end{equation}

Summarizing what we have obtained so far, the five-dimensional metric takes the form
\begin{equation}
 ds^2 =  f [ 2 du (dv + H du) - dy^{\underline{r}} dy^{\underline{r}} ] \,,
\end{equation}
where $H$ is harmonic respect to the metric $f^3 \delta_{\underline{rs}}$
and the scalars $f$ and $\phi^x$ and the v.e.v. of $q^X$ are subject to $h_M = k^X = 0$ and Eq.~(\ref{eq:rdfhi}). Concrete solutions of this equations depend on the value of the constant momentum map $\vec{P}_I$. In any case, we see that, in contrast to the time-like case, in the null case scalar-gravity supersymmetric solutions can have non constant scalars fields $\phi^x$. The presence of the harmonic function $H$ suggests the coupling to external sources.

We may further refine the characterization for the case of constant scalar fields $\phi^x$. As we have already mentioned, this kind of solutions must be critical points of the scalar potential. If $\partial_{\underline{r}} \phi^x = 0$ the Eq.~(\ref{eq:rdfhi}) takes the form
\begin{equation}
 h_I \partial_{\underline{r}} f^{-1/2}  =
   {\textstyle\frac{1}{\sqrt{3}}} g P_I^{\underline{r}} \,,
\label{eq:hidf}
\end{equation}
and its projection to $h^{I}$ yields
\begin{equation}
  \partial_{\underline{r}} f^{-1/2}  = 
   {\textstyle\frac{\sqrt{2}}{\sqrt{3}}} g P^{\underline{r}}  \,.
\label{eq:explicitf}
\end{equation}
This can be easily integrated,
\begin{equation}
 f^{-1/2} = {\textstyle\frac{\sqrt{2}}{\sqrt{3}}} g P^{\underline{r}}  
    y^{\underline{r}} + c \,,
\label{eq:vacuumf}
\end{equation}
where $c$ is a integration constant. On the other hand, the projection of Eq.~(\ref{eq:hidf}) to $h^{\tilde{I}}_x$ yields  $\vec{P}_x  =  0$. Thus we see that the scalar potential receive only the $P^2$ contribution. Following the analysis done in the time-like case, it is easy to see that the first derivatives of the scalar potential vanish.

The presence of the function $f$, whose expression (\ref{eq:vacuumf}) depends on $\vec{P}$, affects the space-time geometry. We may extract further information if we consider two cases separately:

\paragraph{Case 1: $\vec{P} = 0$ \\}
In this case (or in the ungauged limit) the function $f$ becomes constant and it can be absorbed by coordinate rescalings. Thus the metric is
\begin{equation}
 ds^2 =  2 du (dv + H du) - dy^{\underline{r}} dy^{\underline{r}}  \,,
 \hspace{2em}
 \partial_{\underline{r}} \partial_{\underline{r}} H = 0 \,.
\end{equation}
If $H = 0$ we arrive at 5d Minkowski space-time, as we expected since for $\vec{P} = 0$ the scalar potential vanishes.

\paragraph{Case 2: $\vec{P} \neq 0$ \\}
We perform a further coordinate transformation: pick up two constant, unit $SU(2)$ vectors $\vec{e}\,{}^{\underline{i}}$ in such a way that $(\vec{e}\,{}^{\underline{i}}, \vec{e}_P)$ be an orthonormal triad. We define the coordinate system $(z^{\underline{i}},w)$ by
\begin{equation}
  z^{\underline{i}} = e^{\underline{ir}} y^{\underline{r}} \,,
  \hspace{2em}
  w = e_P^{\underline{r}} y^{\underline{r}} 
    + \sqrt{\chi} c \,,
  \hspace{2em}
  \chi \equiv {\textstyle\frac{3}{2}} (gP)^{-2} \,,
\end{equation}
such that $f = \chi w^{-2}$ and $dy^{\underline{r}} dy^{\underline{r}} = dz^{\underline{i}} dz^{\underline{i}} + dw^2$. The five-dimensional metric takes the form
\begin{eqnarray}
 & & ds^2  \ = \  
   \frac{\chi}{w^2} [ 2 du (dv + H du) 
     - dz^{\underline{i}} dz^{\underline{i}} - dw^2 ] \,, 
\\ \nonumber \\
 & & \partial_{\underline{i}} \partial_{\underline{i}} H + (w^{-3} H')' 
   \ = \ 0 \,,
\end{eqnarray}
where the prime stand for the derivative w.r.t $w$. It is evident that if $H = 0$ this metric is $AdS_5$, which is the natural solution since the scalar potential with its sole $P^2$ term yields a negative cosmological constant.


\section{Solutions in $SO(4,1)/SO(4)$}
\label{sec:so41so4}


As we mentioned in the Introduction, in Ref.~\cite{Jong:2006za} an exact supersymmetric solution of $\mathcal{N}=(1,0)$ gauged six dimensional supergravity was found. The solution is a dyonic string with active hyperscalars. In the model used the hyperscalars take values in the $SO(4,1)/SO(4)$ manifold and the $G$-gauge group is $SU(2)$. The model also has one tensor field, being the magnetic/electric charges of the dyonic string defined in terms of it. Using the same $SO(4,1)/SO(4)$ target, two supersymmetric solutions of the ungauged five-dimensional theory coupled to vector- and hypermultiplets were found in Ref.~\cite{Bellorin:2006yr}. These solutions have a point-like naked singularity. In general, the supersymmetric solutions of the 6d and 5d theories, the former classified in Ref.~\cite{Jong:2006za}, have in common that the condition (\ref{eq:rquaternionicmap}) has the same structure in both cases\footnote{In 6d all the supersymmetric solutions fall in the null class. The corresponding transverse space is four-dimensional and can be compared with the base spatial manifold of the time-like class in the 5d theory.}. These results are encouraging to look for more 5d supersymmetric solutions in the $SO(4,1)/SO(4)$ model for the hyperscalars, now within the frame of supersymmetric solutions of the gauged theory with general matter couplings, whose characterization we have performed in section \ref{sec:susy}. 

It must be pointed out that the active tensor field of the solution found in Ref.~\cite{Jong:2006za} is a \emph{gauge} or \emph{massless} tensor field, that is, it appears in the Lagrangian only through its exterior derivative $G=dB_{(6)}+\cdots$. Indeed, this tensor field is the combination of two self/anti-selfdual ones that satisfy the usual massless duality relations in 6d, $\star_6 G^{\pm} \sim \pm G^{\pm}$. This 6d tensor field should not be compared with the tensor fields of the five-dimensional theory we have used in this paper because the latter are, as we have already pointed out, \emph{massive} fields.


\subsection{The model}

We start by defining the geometry of the $SO(4,1)/SO(4)$ manifold and the directions for the gauging. Some preliminaries we need have been already shown in Ref.~\cite{Bellorin:2006yr}, we rewrite them here in order to get a self-contained discussion. We use underlined indices, $\underline{X}=1,2,3,4$, as curved indices in $SO(4,1)/SO(4)$ and non underlined indices $X$ as flat indices. Therefore the coordinates of this manifold are denoted by $q^{\underline X}$. The metric is
\begin{equation}
g_{\underline{XY}} = \Lambda^2\delta_{\underline{XY}}\,,
\hspace{10mm}
\Lambda(q^2) = \frac{2\sqrt2}{1 - q^{\underline X} q^{\underline X}} \,.
\label{eq:qkmetric}
\end{equation}
It can be checked that this metric is Einstein, and since it is also conformally flat, it is trivially selfdual. Therefore this metric is a four-dimensional quaternionic-K\"ahler manifold\footnote{The scalar curvature is $6$, as required by the supersymmetry of the supergravity Lagrangian.}. A vierbein for this metric is
\begin{equation}
        E^X \ =\ \Lambda \delta^X{}_{\underline Y}\ dq^{\underline Y}\, ,
        \hspace{1cm}
        E_X \ =\ \Lambda^{-1} \delta_X{}^{\underline Y}
                \frac{\partial}{\partial q^{\underline Y}}\, .
\end{equation} 
The anti-selfdual part of the spin connection is
\begin{equation}
\Omega^{(-)XY} = 
\frac{1}{\sqrt2} \left(q^{[X} E^{Y]}- \frac{1}{2}\epsilon^{XYWZ} 
q^W E^Z\right)\,,
\end{equation}
where $q^X \equiv \delta^X{}_{\underline Y}\ q^{\underline Y}$. In both the coordinate and the Vierbein basis the three complex structures are given by the 't Hooft symbols,
\begin{equation}
 J^r{}_X{}^Y = 
\delta_X{}^{\underline{W}} J^r{}_{\underline{W}}{}^{\underline{Z}} 
  \delta_{\underline{Z}}{}^Y 
= \rho^r{}_{XY}
\end{equation}
which are real, constant and anti-selfdual matrices in the $X,Y$ indices. This leads us to establish a simple relation between $\Omega^{(-)XY}$ and the $SU(2)$ connection $\vec{\omega}_X$,
\begin{equation}
 \Omega^{(-)}_{XY}{}^Z  =  
   - \vec{\omega}_X  \cdot \vec{J}_Y{}^Z\,.
\label{eq:relatedtargetconnections}
\end{equation}

The group of isometries of $SO(4,1)/SO(4)$ is $SO(4)$. The corresponding Killing vectors, their $\mathfrak{so}(4)$ algebra and their associated momentum maps are given by
\begin{equation}
\begin{array}{llll}
& k_{r}{}^{\underline X} = 
{\textstyle\frac{1}{2}} \rho^r{}_{\underline{XY}} q^{\underline Y}
\,, \hspace{2em}
& k_{r'}{}^{\underline X} = 
{\textstyle\frac{1}{2}}	 \eta^{r'}{}_{\underline{XY}} q^{\underline Y} \,,
\hspace{2em} &
\\ \\
& P^s_{I=r} = {\textstyle\frac{1}{2\sqrt{2}}} \Lambda \delta_{rs} \,,
& P^s_{I=r'} = 0 \,, &
\\ \\
& [k_r , k_s]^{\underline{X}} = -\epsilon_{rst} k_t{}^{\underline{X}} \,,
& [k_{r'} , k_{s'}]^{\underline{X}} = -\epsilon_{r's't'} k_{t'}{}^{\underline{X}} \,,
& [k_r , k_{r'}]^{\underline{X}} = 0 \,.
 \end{array}
\label{eq:killingtarget}
\end{equation}

Now we define the setting for the $G$-group. If a subgroup of the $G$-group acts \emph{non-trivially} on the hyperscalars, then  these must transform in a representation given by a subset of the Killing vectors (\ref{eq:killingtarget}). On the other hand, if a sector of the $G$-group leaves the hyperscalars invariant, then one demand simply that the corresponding Killing vectors (of $\mathcal{M}_{QK}$) vanish, as well as their associated momentum maps. In our model the subgroup of the $G$-group that acts non-trivially on the hyperscalars is $SU(2)$, to which we choose the first set of Killing vectors, $k_r{}^{\underline{X}}$, together with their corresponding momentum maps. 

The $G$-group have to be also a subgroup of the isometry group of the very special manifold $\mathcal{M}_{VS}$. This condition is established in terms of the constraint (\ref{eq:Cgauge}), which we rewrite here for convenience
\begin{equation}
t_{I(\tilde J}{}^{\tilde F} C_{\tilde K\tilde L)\tilde F} = 0 \,.
\label{gaugingveryspecialgeometry}
\end{equation}
Let us start by analyzing this constraint without tensor multiplets. It is straightforward to see that it avoids the gauging of a pure $\mathfrak{su}(2)$ algebra. That is, if one starts by coupling the pure supergravity to two vector multiplets, such that the range of the index $I = r$ is three, and additionally declares that the $G$-group is $SU(2)$, then one faces with the fact that there are not solutions of the constraint
\begin{equation}
 \epsilon_{rs(t} C_{uv)s} = 0
\end{equation}
for non-vanishing $C_{rst}$. In view of this, we use three vector multiplets in our model, such that the total number of vector fields is four, and declare that the fourth direction of the $G$-algebra is Abelian,
that is $G = SU(2) \times U(1)$. One may check that the constraint (\ref{gaugingveryspecialgeometry}) is satisfied by the model
\begin{equation}
 C_{444} = 1 \,,
\hspace*{2em}
C_{4rs} = -{\textstyle\frac{1}{2}} \delta_{rs} \,,
\label{eq:yangmillssector}
\end{equation}
and zero the rest of components of $C_{IJK}$, being $f_{rs}{}^t = \epsilon_{rst}$ the only non-vanishing structure constants. The numerical values has been chosen by convenience. In order to keep the model as close as possible to the previous analyses in the $SO(4,1)/SO(4)$ model for the hyperscalars, we require that these are invariant under the action of the $U(1)$ sector, that is, $k_4{}^{\underline{X}} = \vec{P}_4 = 0$.

Now we add up tensor multiplets, turning on two of them, $n_T = 2$. For concreteness, the matrix $\Omega_{MN}$ can be taken as $\Omega_{12} = 1$.
As we mentioned in section \ref{sec:theory}, not all the symbols $t_{I\tilde{J}}{}^M$ can be zero. Since in our model $ G = SU(2) \times U(1) $, evidently the simplest choice is to put the tensors fields charged respect the $U(1)$ sector, forming a real doublet of it, but invariant under the $SU(2)$ sector. In addition, we may require that tensor and vector fields do not mix under the action of the $G$-group. These three requirements are meet with the representation
\begin{equation}
 t_{rs}{}^t = \epsilon_{rst} \,,
 \hspace{2em}
 t_{4N}{}^M =  {\textstyle\frac{2}{\sqrt{3}}} \Omega^{MN}  
\end{equation}
and zero the rest of components. Due to their block-diagonal form, it is easy to see that the matrices $t_I$ realize the $\mathfrak{su}(2)\times\mathfrak{u}(1)$ algebra, as required in Eq.~(\ref{eq:galgebra}), and satisfy the constraint (\ref{eq:reducible}). Putting these matrices in the formula (\ref{eq:CM}) we find that the only non-vanishing components of $C_{\tilde{I}\tilde{J}M}$ are
\begin{equation} 
 C_{4MN} = - {\textstyle\frac{1}{2}} \delta_{MN} \,.
\label{eq:tensorsector}
\end{equation}
The values of $C_{\tilde{I}\tilde{J}\tilde{K}}$ given in Eqs.~(\ref{eq:yangmillssector}) and (\ref{eq:tensorsector}) define the very special manifold $\mathcal{M}_{VS}$. Having the matrices $t_I$ and $C_{\tilde{I}\tilde{J}\tilde{K}}$, it can be checked by direct computations that this model satisfies the constraint (\ref{gaugingveryspecialgeometry}).



\subsection{The solutions}

We look for the solutions in the time-like class. The scalars fields $\phi^x$ are vanishing, hence the scalars $h^{\tilde{I}}$ and $h^{\tilde{I}}_x$ are nothing but constants. Since they are involved explicitly in the expressions of supersymmetric solutions, it is important to have a concrete set of constant $h^I, h^I_x$ at hand. 
In our model the constraint (\ref{eq:veryspecial}) and its first derivative w.r.t. $\phi^x$ take the form
\begin{eqnarray}
 (h^4)^3 
 -{\textstyle\frac{3}{2}} h^4 h^i h^i & = & 1 \,,
 \\ \nonumber \\
 h^4_x [ (h^4)^2  - {\textstyle\frac{1}{2}} h^i h^i ] 
 - {\textstyle\frac{1}{2}} h^4 h^i h^i_x 
 & = & 0 \,,
\end{eqnarray}
where $h^i=(h^r,h^M)$. A consistent solution to this system is $h^4 = 1$, $h^i = 0$, $h^4_x = 0$ and $h^i_x \neq 0$,  which yields the simplest version for the scalar metric, $a_{\tilde{I}\tilde{J}} = \delta_{\tilde{I}\tilde{J}}$. The (constant) induced metric $g_{xy}$ is diagonalized by $h^i_x$, $g_{xy} = h^i_x h^j_y \delta_{ij}$. All the Killing vectors of $\mathcal{M}_{VS}$, which are defined in Eq.~(\ref{eq:killingspecial}), vanish for this solution. The tensor mass matrix defined in Eq.~(\ref{eq:tensormass}) yields $[\mathcal{M}^2]^M{}_N = 4 g \delta^M{}_N$. 

Now we turn our attention to the space-time metric given in Eq.~(\ref{eq:rmetric}). We put $\omega = 0$. We assume that the base manifold is conformally flat, 
\begin{equation}
h_{\underline{mn}}dx^{\underline m} dx^{\underline n}
= \Omega^2 dx^{\underline m} dx^{\underline m}\,,
\hspace{10mm}
\Omega = \Omega(r^2)\, ,
\hspace{10mm}
r^2 \equiv x^{\underline m} x^{\underline m} \; ,
\end{equation}
and hence take the Vierbein on the base manifold to be
\begin{equation}
        v^m = \Omega \delta^m{}_{\underline m}dx^{\underline m}\,,
        \hspace{10mm}
        v_m = \Omega^{-1}\delta_m{}^{\underline m}\partial_{\underline m}\,.
\end{equation}
In this basis we can identify the complex structures of the base manifold with
those of the hypervariety
\begin{equation}
\Phi^r{}_{m}{}^{n}
= \delta_m{}^X J^r{}_X{}^Y \delta_Y{}^n
= \rho^r{}_{mn}\, .
\end{equation}
The anti-selfdual part of the spin connection on the base manifold is
\begin{equation}
\xi^{(-)mn} = 2\frac{\Omega'}{\Omega^2}
\left(
x^{[m} v^{n]} - {\textstyle\frac{1}{2}}\epsilon^{mnpq} x^p v^q
\right)
\end{equation}
where $x^m = \delta^m{}_{\underline m}\ x^{\underline n}$.

Regarding the supersymmetric vector/tensor fields given in Eqs.~(\ref{eq:rmetric}) and (\ref{eq:rFI}), we see that only the Abelian vector field has time-like component,
\begin{equation}
 A^4 = -\sqrt{3} f dt + \hat{A}^4 \,,
\hspace{2em}
A^r = \hat{A}^r \,,
\hspace{2em}
B^M = \hat{B}^M \,.
\end{equation}
These spatial fields are subject to the preserved supersymmetry conditions (\ref{eq:rH+}) and (\ref{eq:rH-}). The former yields $ \hat{F}^{4(+)} = 0$, whereas the $\tilde{I} = (4,M)$ components of the latter yield 
\begin{equation}
 \hat{F}^{4(-)} = 0 \,,
 \hspace{2em}
 \hat{B}^{M(-)} = 0 \,.
\label{eq:selfdualB}
\end{equation}
Thus we conclude that $\hat{A}^4 = 0$, hence the Abelian gauge field is an electrostatic field, $A^4 = -\sqrt{3} f dt$. The $\tilde{I} = r$ components of Eq.~(\ref{eq:rH-}) becomes
\begin{equation}
 \hat{F}^{r(-)}_{\underline{mn}} =
 {\textstyle\frac{1}{2\sqrt{2}}} g f^{-1} \Omega^2 \Lambda \rho^r_{\underline{mn}} \,.
\label{eq:F-}
\end{equation}

Now we analyze the Eq. (\ref{eq:rquaternionicmap}). In our setting this equations takes the form
\begin{equation}
\hat{\mathfrak{D}}_{\underline m} q^{\underline X} =
\left(\delta_{\underline{mY}}\delta_{\underline{nX}}
- \delta_{\underline{mX}}\delta_{\underline{nY}}
-\epsilon_{\underline{mnYX}}\right)
\hat{\mathfrak{D}}_{\underline n} q^{\underline Y}
\end{equation}
whose symmetric and antisymmetric parts give
\begin{eqnarray}
\hat{\mathfrak{D}}_{\underline m}q^{\underline m} & =& 0\, , 
\label{eq:symmquaternionic}
\\ \nonumber \\
\hat{\mathfrak{D}}_{[\underline m} q_{\underline n]} & =&
        -{\textstyle\frac{1}{2}}\epsilon_{\underline{mnpq}}
        \hat{\mathfrak{D}}_{\underline p} q_{\underline q}\,,
\label{eq:antisymmquaternionic}
\end{eqnarray}

\noindent
where $q_{\underline m} = q^{\underline m}$. 

To solve these equations we use the same ansatz of Ref.~\cite{Jong:2006za},
\begin{equation}
q^{\underline m} = x^{\underline{m}} Q \,,
\hspace*{2em}
\hat{A}^r{}_{\underline{m}} = \rho^r{}_{\underline{mn}} x^{\underline{n}} A \,,
\label{eq:ansatz}
\end{equation}
where
\begin{equation}
 Q = Q(r^2) \,,
\hspace{2em}
A = A(r^2) \,.
\end{equation}
Eq.~(\ref{eq:antisymmquaternionic}) is automatically solved by this ansatz, whereas the Eq.~(\ref{eq:symmquaternionic}) is solved if
\begin{equation}
r^2 Q' + \left(2 + {\textstyle\frac{3}{4}} g r^2 A \right) Q = 0 \,.
\label{eq:eq1}
\end{equation}
This equation fixes one of the scalars $A$ or $Q$ in terms of the other one.

The field strength of the gauge field given in Eq.~(\ref{eq:ansatz}) is
\begin{eqnarray}
\hat{F}^{r(+)}_{\underline{mn}} 
& = &
2 ( \rho^r{}_{\underline{p} [\underline{m}} x_{\underline{n}]} x^{\underline{p}}
  +{\textstyle\frac{1}{4}} \rho^r{}_{\underline{mn}} r^2 )
(2A' - g A^2) \,, 
\label{eq:F+}
\\ \nonumber \\
\hat{F}^{r(-)}_{\underline{mn}} 
& = &
-\rho^r{}_{\underline{mn}} ( r^2 A'+ 2A + {\textstyle\frac{1}{2}} g r^2 A^2 )\,.
\label{eq:Fansatz}
\end{eqnarray}
By contrasting last expression with Eq.~(\ref{eq:F-}) we obtain the condition
\begin{equation}
 r^2 A' + 2 A + {\textstyle\frac{1}{2}} g r^2 A^2 =
- {\textstyle\frac{1}{2\sqrt{2}}} g f^{-1} \Omega^2 \Lambda  \,.
\label{eq:eq2}
\end{equation}

The last condition required by preserved supersymmetry is given in Eq.~(\ref{eq:rspin}), whose both sides can be evaluated on the configuration as we have it. This yields
\begin{equation}
 \frac{\Omega'}{\Omega} =
\frac{2 Q^2 + g A}{2(1-r^2Q^2)} \,.
\label{eq:eq3}
\end{equation}
This equation was already found in the six dimensional theory in Ref.~\cite{Jong:2006za}, being $\Omega$ the conformal factor of the transverse metric in that paper. If we solve Eq.~(\ref{eq:eq1}) for $A$ and substitute it in Eq.~(\ref{eq:eq3}), the resulting equation can be integrated, as was done in Ref.~\cite{Jong:2006za}, yielding
\begin{equation}
\Omega =
\frac{1}{r^2} \left(\frac{1 - r^2 Q^2}{r^2 Q^2} \right)^{1/3} \,,
\label{eq:Omega}
\end{equation}
where we have put an irrelevant multiplicative integration constant equal to one.

Having obtained $\Omega$, Eq.~(\ref{eq:eq2}) can be used to obtain an expression for $f$,
\begin{equation}
g f^{-1} =
 - 2\sqrt{2} (\Omega^2 \Lambda)^{-1} 
   (r^2 A' + 2A + {\textstyle\frac{1}{2}} g r^2 A^2) \,.
\label{eq:f}
\end{equation}

So far, we have analyzed the conditions imposed by preserved supersymmetry on the configuration. Now it is time to check the equations of motion, which are given in Eqs.~(\ref{eq:rmaxwell}) and (\ref{eq:rdb}). One can check that the configuration, as we have it, solves the equation (\ref{eq:rmaxwell}) except for the $I=4$ component, which yields
\begin{equation}
 \hat{\nabla}^2 f^{-1} 
- 2 \Omega^{-4} (r^2 A' + A)(A + g r^2 A^2)
+ {\textstyle\frac{1}{12}} \Omega^{-4} \hat{B}^2
= 0 \,,
\label{eq:eq4}
\end{equation}
where $\hat{B}^2 = \hat{B}^M_{\underline{mn}} \hat{B}^M_{\underline{mn}}$ and we have used the fact that $\hat{B}^M$ are selfdual.

The tensor Eq.~(\ref{eq:dB}) evaluated on the configuration becomes (for $g \neq 0$)
\begin{equation}
\hat{\mathfrak{D}} \hat{B}^M = d\hat{B}^M = 0 \,.
\end{equation}
Therefore the spatial tensors $\hat{B}^M$ are exact two-forms, $\hat{B}^M = d\hat{C}^M$, with $\hat{C}^M$ satisfying the selfduality condition $\hat{\star} d \hat{C}^M = d \hat{C}^M$. When evaluated in components, it turns out that this condition is independent of the conformal factor $\Omega$,
\begin{equation}
 {\textstyle\frac{1}{2}} \epsilon_{\underline{mnpq}} \partial_{\underline{p}}
 \hat{C}_{\underline{q}} = \partial_{[\underline{m}} \hat{C}_{\underline{n]}} \,.
\label{eq:selfdual}
\end{equation}
At first sight, one may feel uncomfortable with tensor fields that are derived from one-forms. However, we should take into account that, after all, $B^M = d\hat{C}^M$ leads to a non-trivial solution of the equation of motion (\ref{eq:EM}). That is, although $\hat{\mathfrak{D}}\hat{B}^M = 0$, we have that $\mathfrak{D} B^M \neq 0$ hence the two terms of the Eq.~(\ref{eq:EM}) are non-zero but cancel mutually. It is this equation what distinguish between vector and tensor fields. We also recall that tensor fields transform in a non-adjoint representation of the $G$-group. In our model this means that $B^M$ are charged under $U(1)$. Therefore, there is no way to identify the one-forms $\hat{C}^M$ with some physical vector gauge fields. 


At this stage the solution is build from three elements: the scalars $A$ and $Q$ and the one-forms $\hat{C}^M$. These variables must be chosen in such a way that the conditions given in Eqs.~(\ref{eq:eq1}), (\ref{eq:Omega}), (\ref{eq:f}), (\ref{eq:eq4}) and (\ref{eq:selfdual}) are satisfied. We have found three classes of solutions of these conditions: two classes in the gauged theory, one with active tensors fields and the other one without tensor fields, and one class in the ungauged theory.


\subsubsection{Solutions of the gauged theory}
There is a route to find solutions of the Eqs.~(\ref{eq:eq4}) and (\ref{eq:selfdual}). First off all, note that the expression $r^2 A' + A$ vanishes for a square-inverse dependence of $A$,
\begin{equation}
 A = \frac{a}{r^2} \,,
\label{eq:Aansatz}
\end{equation}
where $a$, as well as $b$, $c$ and $d$ for future reference, are integration constants. We take this ansatz as the starting point, such that the second term of the Eq.~(\ref{eq:eq4}) vanishes, and consequently we require that the first and last terms be non-zero but cancel mutually in order to get non-trivial tensor fields. 

We may substitute the ansatz (\ref{eq:Aansatz}) for $A$ in Eq.~(\ref{eq:eq1}) obtaining a differential equation for $Q$,
\begin{equation}
r^2 Q' + k Q = 0 \,, 
\hspace{2em}
k \equiv {\textstyle\frac{1}{4}} ( 8 + 3 g a ) \,,
\end{equation}
whose solution, for any value of $k$, is 
\begin{equation}
 Q = \frac{b}{r^{2k}} \,.
\label{eq:Qansatz}
\end{equation}
Having $Q$, one can compute straightforwardly $\Omega$ and $f^{-1}$ from Eqs.~(\ref{eq:Omega}) and (\ref{eq:f}). Doing so we obtain
\begin{eqnarray}
 \Omega & = & \frac{1}{b^{2/3}} \frac{(r^{2(2k-1)} - b^2)^{1/3}}{r^2} \,, 
\\ \nonumber \\
 f^{-1} & = & k' \frac{(r^{2(2k-1)} - b^2 )^{1/3}}{r^{4(k-1)}} \,,
\hspace*{2em}
 k' \equiv - \frac{ 8 b^{4/3}}{9 g^2} (k - 2) (k - 1/2) \,.
\end{eqnarray}
Regularity of the configuration imposes $b \neq 0$, hence hyperscalars can not be smoothly turned off from this class of configurations (already known from Eq.~(\ref{eq:Omega})). 
The Laplacian of $f^{-1}$ w.r.t. the metric $h_{\underline{mn}}$ yields a somewhat simple expression,	
\begin{equation}
 \hat{\nabla}^2 f^{-1} =
 \frac{k'' \Omega^{-4}}{r^{2(2k+1)}} \,,
\hspace{2em}
k'' \equiv
\frac{128 b^2}{9 g^2} (k - 2) (k - 1) (k - 1/2)^2 \,.
\label{eq:laplacianf}
\end{equation}
Substituting these expressions into the Maxwell Eq.~(\ref{eq:eq4}) we obtain
\begin{equation}
 \hat{B}^2 = 
 - \frac{12 k''}{r^{2(2k+1)}} \,.
\label{eq:b2}
\end{equation}
In order to solve this equation with non-vanishing tensor fields, we are forced to demand $k'' < 0$, hence the range of admissible values of the exponent $k$ is
\begin{equation}
  1 < k < 2 \,.
\label{eq:krange}
\end{equation}
On the other hand, if we look for solutions with vanishing tensor fields, then $k'' = 0$ and this is in principle possible for three values of the exponent $k$, $k = 1/2, 1,2$. However, the values $1/2$ and $2$ yield a vanishing $k'$, hence a singular $f$. Therefore, the only exponent valid for vanishing tensor fields is $k = 1$. We remark that, at this level, all the functional form of the supersymmetric solution depends only the one-forms $\hat{C}^M$, they must solve the selfduality condition (\ref{eq:selfdual}) and also the Maxwell Eq.~(\ref{eq:b2}) for an appropriated exponent $k$.


\paragraph{Solution with active tensors fields\\}
The expressions for the vector fields $\hat{A}^r$ given in Eqs.~(\ref{eq:ansatz}), (\ref{eq:F+}) and (\ref{eq:Fansatz}) suggest an ansatz to solve the Eq.~(\ref{eq:selfdual}), it is 
\begin{equation}
 \hat{C}^M_{\underline{m}} = \alpha^M_r \rho^r_{\underline{mn}} x^{\underline{n}} C \,,
 \hspace*{2em}
 C = C(r^2) \,,
\end{equation}
and $\alpha^M_r$ is a set of constants. The corresponding self/anti-selfdual parts of $\hat{B}^M = d \hat{C}^M$ are
\begin{eqnarray}
 \hat{B}^{M(+)}_{\underline{mn}} & = &
    4 \alpha^M_ r (\rho^r{}_{\underline{p} [ \underline{m}} x_{\underline{n} ]} 
     x^{\underline{p}} 
      + {\textstyle\frac{1}{4}} \rho^r_{\underline{mn}} r^2 ) C' \,,
\\ \nonumber \\
 \hat{B}^{M(-)}_{\underline{mn}} & = &
   - \alpha^M_r \rho^r_{\underline{mn}} ( r^2 C' + 2 C )
    \,.
\end{eqnarray}
Therefore, in order to have non-trivial and selfdual spatial tensors $\hat{B}^M$, we are forced to impose
\begin{equation}
 r^2 C' + 2 C = 0 \,,
\end{equation}
whose solution is
\begin{equation}
 C = \frac{1}{r^4} \,,
\end{equation}
where any multiplicative integration constant can be absorbed in the value of $\alpha^M_r$. Thus, we have that the selfdual spatial tensor fields are given by
\begin{eqnarray}
 \hat{C}^M_{\underline{m}} & = &
  \alpha^M_r \rho^M_{\underline{mn}} x^{\underline{n}} r^{-4} \,,
\\ \nonumber \\
 \hat{B}^M_{\underline{mn}} & = &
  - 8 \alpha^M_r r^{-6} (\rho^r{}_{\underline{p} [ \underline{m}} x_{\underline{n} ]} 
        x^{\underline{p}} 
      + {\textstyle\frac{1}{4}} \rho^r_{\underline{mn}} r^2 ) \,.
\end{eqnarray}
The square of these selfdual spatial tensor fields yields
\begin{equation}
 \hat{B}^2 = \frac{16 \alpha^2}{r^8} \,,
 \hspace*{2em}
 \alpha^2 \equiv \alpha^M_r \alpha^M_r \,.
\end{equation}
The final and crucial test for our ansatz is that this square must be a solution of Eq.~(\ref{eq:b2}). This holds only for 
\begin{equation}
 k = \frac{3}{2} \,,
\end{equation}
which is in the range of admissible values of the exponent $k$ given in the inequalities (\ref{eq:krange}). Therefore, the Maxwell Eq.~(\ref{eq:b2}) is solved for $k=3/2$ if $\alpha^2 = 8 b^2 /(3 g^2)$. 

In summary, the five-dimensional solution, in Cartesian coordinates, is\footnote{To write this solution in a simple form, we have put $b=1$, $g=1/3$, rescaled the constants $\alpha^M_r$ and rescaled the time $t \rightarrow 4 t$.} 
\begin{equation}
 \begin{array}{rclrcl}
  ds^2 & = & 
   r^{4} ( r^4 - 1 )^{-2/3}  dt^2 
   - 4 r^{-6} ( r^4 - 1 )  dx^{\underline{m}} dx^{\underline{m}} \,, & & &
\\ \\
  q^{\underline{m}} & = & r^{-3} x^{\underline{m}} \,, &
  \phi^x & = & 0   \,, 
\\ \\
  A^r & = &  2 \rho^r_{\underline{mn}} x^{\underline{m}} r^{-2} dx^{\underline{n}} \,, & 
  A^4 & = & - \sqrt{3} r^{2} ( r^4 - 1 )^{-1/3} dt \,,
\\ \\
  F^r & = & 
   -\frac{16}{3} r^{-4} ( \rho^r{}_{\underline{p} [\underline{m}} 
      x_{\underline{n}]} x^{\underline{p}}
        + {\textstyle\frac{1}{2}} \rho^r{}_{\underline{mn}} r^2 ) 
           dx^{\underline{m}} \wedge dx^{\underline{n}} \,, & 
  F^4 & = & 
   -{\displaystyle\frac{ 2 \sqrt{3} (\frac{1}{3} r^4 - 1)}{ (r^4 - 1)^{4/3}}} x^{\underline{m}} 
      dx^{\underline{m}} \wedge dt \,, 
\\ \\
  B^M & = & 
  - 8\sqrt{6} \alpha^M_r r^{-6} (\rho^r{}_{\underline{p} [ \underline{m}} 
  x_{\underline{n} ]} x^{\underline{p}} 
    + {\textstyle\frac{1}{4}} \rho^r_{\underline{mn}} r^2 )
  dx^{\underline{m}} \wedge dx^{\underline{n}} \,. \hspace*{-5mm} & & &
\end{array}
\label{eq:sol1}
\end{equation}
where $\alpha^M_r$ are six constants normalized by $\alpha^2 = 1$. This normalization condition is the only constraint on the constants $\alpha^M_r$, which leaves us the freedom to turn off one the tensor fields, for example by setting $\alpha^{M=2}_r = 0$. Notice that no further active fields can be smoothly turned off in this solution.

Regularity of the metric restricts the domain of the solution to $1 < r < \infty$. To determine the coordinate-dependence of the divergences as $r$ approach to $1$ and $\infty$, we compute the Ricci scalar, it yields 
\begin{equation}
 R = 
  \frac{r^4 (23 r^8 + 114 r^4 - 45)}{ 18 (r^4 - 1)^3} \,.
\end{equation}
The Ricci scalar diverges as $r \rightarrow 1$ whereas it is completely regular at $r \rightarrow \infty$. Therefore, we conclude that this metric has  a physical singularity and a horizon located, in our coordinate system, at $r = 1$ and $r \rightarrow \infty$ respectively. The regularity of the metric at the horizon becomes more evident if we shift the metric in Eq.~(\ref{eq:sol1}) to spherical coordinates and then perform the coordinate transformation $\tilde{r} = r^{-2/3}$. This leads us to the metric
\begin{equation}
 ds^2 =
 \tilde{r}^{-2} \left[ 
  ( 1 - \tilde{r}^{6} )^{-2/3} dt^2
  - ( 1 - \tilde{r}^{6} ) ( 9 d\tilde{r}^2 + 4 \tilde{r}^2 d\Omega_{(3)}^2 ) \right] \,,
\end{equation}
which, as $\tilde{r} \rightarrow 0$ and after a further rescaling of the time, takes the form
\begin{equation}
 ds^2 =
  9 \tilde{r}^{-2} ( dt^2 - d\tilde{r}^2 ) - 4 d\Omega_{(3)}^2  \,.
\label{eq:ads2s3}
\end{equation}
Therefore, in the near horizon limit the metric approach to $AdS_2 \times S^3$.
We notice that in the domain $0 < \tilde{r} < 1$ the conformal factor of the $\mathcal{M}_{QK}$ metric,
\begin{equation}
 \Lambda =
  2\sqrt{2} (1 - \tilde{r}^{6})^{-1} \,,
\end{equation}
is regular. Finally, we point out that the Cartesian components of the tensor fields $B^ M$ given in Eqs.~(\ref{eq:sol1}) decay asymptotically as $r^{-4}$, faster than any of the field strengths $F^ I$ of the vector fields. This is the expected behavior for massless/massive fields.


\paragraph{Solution without tensors fields \\}
We start by setting the tensor fields equal to zero, $\hat{C}^M = 0$, which is the most simple solution of Eq.~(\ref{eq:selfdual}). As we have already pointed out, only the exponent
\begin{equation}
  k = 1 
\end{equation}
is admissible to solve the Maxwell Eq.~(\ref{eq:b2}). The five-dimensional solution corresponding to this exponent is
\begin{equation}
 \begin{array}{rclrcl}
  ds^2 & = & 
   ( r^2 - 1 )^{-2/3} dt^2 
    - 4 r^{-4}( r^2 - 1 ) dx^{\underline{m}} dx^{\underline{m}} \,,
\\ \\
  q^{\underline{m}} & = &  r^{-2} x^{\underline{m}} \,, &
  \phi^x & = & 0 \,, 
\\ \\
  A^r & = &  4 \rho^r_{\underline{mn}} x^{\underline{m}} r^{-2} dx^{\underline{n}} \,, & 
  A^4 & = & - \sqrt{3} ( r^2 - 1 )^{-1/3} dt \,,
\\ \\
  F^r & = & 
    -\frac{40}{3} r^{-4} ( \rho^r{}_{\underline{p} [\underline{m}} 
      x_{\underline{n}]} x^{\underline{p}}
        + {\textstyle\frac{1}{2}} \rho^r{}_{\underline{mn}} r^2 ) 
           dx^{\underline{m}} \wedge dx^{\underline{n}} \,, \hspace*{5mm} & 
  F^4 & = & 
   - \frac{2}{\sqrt{3}}(r^2 - 1)^{-4/3} x^{\underline{m}} 
      dx^{\underline{m}} \wedge dt \,, 
\\ \\
  B^M & = & 0 \,. & & &
\end{array}
\end{equation}
Notice that, according to Eq.~(\ref{eq:laplacianf}), the function $f^{-1} = ( r^2 - 1 )^{1/3}$ is harmonic respect to the spatial metric\footnote{Indeed, any spherically symmetric harmonic function $H(r^2)$ must be of the form $H' \sim r^{-4} \Omega^{-2}$, for any $\Omega$.} $h_{\underline{mn}}$, whose conformal factor is $\Omega = r^{-2} (r^2 -1)^{1/3}$. None of the active fields can be smoothly turned off from this solution.

Again, regularity of the metric demands $1 < r < \infty$. The corresponding Ricci scalar is
\begin{equation}
 R = 
  \frac{23 r^6}{18 (r^2 - 1)^3} \,.
\end{equation}
As in the previous case, this metric has a physical singularity located at $r = 1$ and a horizon at $r \rightarrow \infty$. After performing the coordinate transformation $\tilde{r} = r^{2/3}$, this metric becomes
\begin{equation}
 ds^2 = 
 \tilde{r}^{-2} \left[ 
  ( 1 - \tilde{r}^{-3} )^{-2/3} dt^2
  - ( 1 - \tilde{r}^{-3} ) ( 9 d\tilde{r}^2 + 4 \tilde{r}^2 d\Omega_{(3)}^2 ) \right] \,,
\end{equation}
which in the near horizon limit $\tilde{r} \rightarrow \infty$ takes the $AdS_2\times S^3$ form (\ref{eq:ads2s3}).
The conformal factor of $\mathcal{M}_{QK}$,
\begin{equation}
 \Lambda = 
  2\sqrt{2} ( 1 - \tilde{r}^{-3} )^{-1} \,,
\end{equation}
is regular in $1 < \tilde{r} < \infty$.


\subsubsection{Solution of the ungauged theory}
It is interesting to analyze the case of the ungauged theory ($g=0$), which, first of all, requires the decoupling of the tensor fields. In this case the model reduces to a ungauged $\mathcal{N}=1$, $d=5$ Supergravity coupled to one hypermultiplet and three vector multiplets, being the four $A^I$ fields Abelian gauge fields.

Notice that now the Eq.~(\ref{eq:f}) does not fix the scalar $f$, it is instead an equation for $A$. Indeed, this and Eq.~(\ref{eq:eq1}) state that $A$ and $Q$ satisfy that same differential equation,
\begin{equation}
 r^2 Q' + 2 Q = r^2 A' + 2 A = 0 \,,
\end{equation}
whose solutions are
\begin{equation}
 A = \frac{a}{r^4} \,,
\hspace*{2em}
 Q = \frac{b}{r^4} \,.
\end{equation}
$\Omega$, computed from Eq.~(\ref{eq:Omega}), becomes
\begin{equation}
 \Omega = \frac{1}{b^{2/3}} \left( 1- \frac{b^2}{r^{6}}\right)^{1/3} \,.
\end{equation}

Evaluating the Maxwell Eq.~(\ref{eq:eq4}) we obtain
\begin{equation}
 \hat{\nabla}^2 f^{-1} 
 + \frac{2 a^2 \Omega^{-4}}{r^{8}} = 0 \,,
\end{equation}
and it is now regarded as an equation for $f$. If we assume that $f^{-1}$ is spherically symmetric, then this equation becomes
\begin{equation}
 r^2 [\Omega^2 (f^{-1})']' + 2 \Omega^2 (f^{-1})' = -\frac{a^2}{2 r^{8}} \,,
\end{equation}
and its solution for $\Omega^2 (f^{-1})'$ is
\begin{equation}
 \Omega^2 (f^{-1})' = 
 \frac{a^2}{4 r^{8}} + \frac{c}{r^4} \,.
\end{equation} 
We can find the integral of this expression, which we write here for the special value $b=1$ for simplicity,
\begin{equation}
 f^{-1} = 
 {\textstyle\frac{1}{4}} a^2 \Omega + H \,,
\end{equation}
where
\begin{equation}
 H = 
   c \, \frac{{}_2F_1 {\textstyle(\frac{1}{3},\frac{2}{3};\frac{4}{3};r^{-6})}}{r^2}
    + d
\end{equation}
is a spherically symmetric harmonic function corresponding to the conformal factor $\Omega = ( 1 - r^{-6} )^{1/3}$ (we have inverted the sign of the arbitrary integration constant $c$).

The whole five-dimensional solution is
\begin{equation}
 \begin{array}{rclrcl}
  ds^2 & = & 
   \left( {\textstyle\frac{1}{4}} a^2 \Omega + H\right)^{-2} dt^2   
    - \Omega^2 \left( {\textstyle\frac{1}{4}} a^2 \Omega
        + H\right) dx^{\underline{m}} dx^{\underline{m}} \,, \hspace*{1em} & & & 
\\ \\
  H & = & 
    c \, r^{-2} \, {}_2F_1 {\textstyle\left(\frac{1}{3},\frac{2}{3};\frac{4}{3};r^{-6}\right)}
    + d   \,, &
  \Omega & = &
    ( 1 - r^{-6} )^{1/3} \,,
\\ \\
  q^{\underline{m}} & = & r^{-4} x^{\underline{m}} \,, &
  \phi^x & = & 0  \,, 
\\ \\
  A^r & = &  a \rho^r_{\underline{mn}} x^{\underline{m}} r^{-4} dx^{\underline{n}} \,, & 
  A^4 & = & - \sqrt{3} \left( {\textstyle\frac{1}{4}} a^2 \Omega + H\right)^{-1} dt \,.
\end{array}
\end{equation}
As for the gauged solutions, the domain of this solution is restricted to $ 1 < r < \infty $ and also $c,d \geq 0$. To analyze the character of the divergences we first put $c = 0$. The resulting metric has a divergence only at $r = 1$, which can be seen by direct computation of the Ricci scalar that correspond to a physical singularity. This singularity is still present in the general solution ($c \neq 0$) since the hypergeometric function ${}_2F_1{\textstyle\left(\frac{1}{3},\frac{2}{3};\frac{4}{3};r^{-6}\right)}$ is convergent, positive definite and monotonically decreasing in the interval $ 1 < r < \infty$. Thus we conclude that the above metric has a naked singularity at $r=1$. 

In contrast to the previous solution, in this solution there are three free parameters, $a$, $c$ and $d$, that can smoothly be brought to zero. The limiting value $a = 0$ leads to the two solutions found in Ref.~\cite{Bellorin:2006yr}, one with $c = 0$ and the other one with $c \neq 0$.

\section{Conclusions}
\label{sec-conclusions}

We have performed the complete characterization of supersymmetric configurations and solutions for the gauged $\mathcal{N}=1$, $d=5$ Supergravity coupled to an arbitrary number of vector-, tensor- and hypermultiplets. This is the most general supergravity theory known with eight supercharges in five dimensions.

As was found in previous analyses without tensor multiplets, Refs.~\cite{Bellorin:2006yr,Bellorin:2007yp}, the spatial spin connection, both in the time-like and null cases, is embedded into other gauge connections given by the quaternionic K\"ahler geometry and the vector fields. The activation of tensor fields do not deform these embeddings. However, tensor fields give new contributions to the supersymmetric Maxwell equation, which in general leads to deformations of the base spatial metrics. Moreover, the presence of tensor fields adds one more equation to be solved by supersymmetric solutions, the self-duality condition for supersymmetric tensor fields.

The conditions for unbroken supersymmetry we have found can be used to study concrete class of supersymmetric solutions. In particular, we have used them to analyze scalar-gravity solutions, which we define by the condition of vector and tensor fields be vanishing and the hyperscalars get a constant value. For these configurations we have found that in the time-like case also the scalars $\phi^x$ must be constant but in the null case they can have a non-trivial expression. For the cases of constant $\phi^x$ we have classified the metric according to the vanishing or not of the constant parameter $\vec{P} \sim h^I \vec{P}_I$, being zero the other contributions to the scalar potential. In the time-like case we have found that if $\vec{P} = 0$ then the base spatial manifold is hyperk\"ahler and the local one-form $\omega$ is vanishing, whereas if $\vec{P}\neq 0$ the base manifold is K\"ahler and $\omega$ is non-vanishing. In the null case the spatial transverse metric is conformally flat and we have seen the presence of a harmonic function, which suggests the direct coupling to external sources. Among the null class, one gets Minkowski or $AdS_5$ if $\vec{P} = 0$ or $\vec{P} \neq 0$ respectively.

Since we have determined the behavior of the several variables that define a scalar-gravity supersymmetric solution, these result can be used to search more general solutions with asymptotics corresponding to pure gravity solutions, for example $AdS_5$. As it is well known, asymptotically $AdS_5$ supersymmetric solutions are of great interest due to their role in the brane-world scenarios and in the AdS/CFT correspondence.

We have also used the conditions for supersymmetric solutions to find new exact solutions with active hyperscalars, vector and tensor fields, which is a further evidence of the usefulness of the characterization program. To find the solutions we have chosen the $SO(4,1)/SO(4)$ manifold as the target for the hyperscalars. Two of the solutions we have found belong to the gauged theory with $SU(2) \times U(1)$ as the gauge group, one of them having active, massive and spatially self-dual tensor fields charged under the $U(1)$ sector and decaying faster than the massless vector fields. These two solutions are black holes (spherically symmetric) with physical singularities covered by the horizons and with $AdS_2 \times S^3$ as the near horizon geometry. The third solution we have found belongs to the ungauged theory. It has a naked physical singularity and has the solutions found in Ref~\cite{Bellorin:2006yr} as limiting cases. It must be pointed out that solutions with naked singularities could be excluded as genuine supersymmetric solutions if a criterion as the one proposed in Ref.~\cite{Bellorin:2006xr}, which consists on requiring unbroken supersymmetry even at the location of the singularity, is used.


\section*{Acknowledgments}
This work has been supported in part by the Universidad Sim\'on Bol\'{\i}var grant S1-IN-CB-002-08.




\end{document}